\documentclass[aps,prl,reprint,showpacs,superscriptaddress]{revtex4-1}

\pdfoutput=1
\usepackage{color}
\usepackage{graphicx}
\usepackage{amsmath}
\usepackage{amssymb}
\usepackage{dsfont}

\newcommand*{\twobytwo}[4]{\begin{pmatrix} #1 & #2 \\ #3 & #4\end{pmatrix}}
\newcommand*{\ket}[1]{|{#1}\rangle}

\begin{document}
\title{Lamb shift enhancement and detection in strongly driven superconducting circuits}
\date{\today}
\author{Vera Gramich}
\email{vera.gramich@uni-ulm.de}
\affiliation{Institut f\"{u}r Theoretische Physik, Universit\"{a}t Ulm,
  Albert-Einstein-Allee 11, 89069 Ulm,
  Germany}
\affiliation{Low Temperature Laboratory (OVLL) - Aalto University School of Science, P.O. Box 13500, 00076 Aalto, Finland}
\author{Simone Gasparinetti}
\affiliation{Low Temperature Laboratory (OVLL) - Aalto University School of Science, P.O. Box 13500, 00076 Aalto, Finland}
  \author{Paolo Solinas}
  \affiliation{Low Temperature Laboratory (OVLL) - Aalto University School of Science, P.O. Box 13500, 00076 Aalto, Finland}
\affiliation{SPIN-CNR - Via Dodecaneso 33, 16146 Genova, Italy}
\author{Joachim Ankerhold}
\affiliation{Institut f\"{u}r Theoretische Physik, Universit\"{a}t Ulm,
  Albert-Einstein-Allee 11, 89069 Ulm,
  Germany}

\begin{abstract}
It is shown that strong driving of a quantum system substantially enhances the Lamb shift induced by broadband reservoirs which are typical for solid-state devices. By varying drive parameters the impact of environmental vacuum fluctuations with continuous spectral distribution onto system observables can be tuned in a distinctive way. This provides experimentally feasible measurement schemes for the Lamb shift in superconducting circuits based on Cooper pair boxes, where it can be detected either in shifted dressed transition frequencies or in pumped charge currents.
\end{abstract}


\maketitle
\paragraph{Introduction.-}
Quantum fluctuations of the electromagnetic vacuum affect atomic spectra \cite{Lamb:1947}, a phenomenon termed Lamb shift (LS)  which has triggered the development of modern quantum electrodynamics (QED). Experimentally, cavity QED \cite{Heinzen:1987,Brune:1994, Marrocco:1998} has opened the door for an unprecedented level of accuracy in the manipulation and measurement of atomic quantum states \cite{Haroche:2006}. Its most recent realization is circuit QED \cite{Wallraff:2004, Schuster:2007, Blais:2007} based on solid-state architectures, for example, consisting of a superconducting Cooper pair box (CPB) embedded in a superconducting waveguide resonator.
Circuit QED has been able to reproduce several quantum optics experiments \cite{Schuster:2005, Fink:2008, Fink:2009}, with advantages in terms of design, fabrication and scalability giving also access to parameter ranges currently unreachable in optical setups \cite{Schuster:2005, Fink:2008, Fink:2009}.

Cavity and circuit QED are based on strongly modified density of states of the electromagnetic environment seen by the atom compared to a continuum.  This way, the LS has recently also been detected in a circuit QED setup \cite{Fragner:2008} in form of zero-point fluctuations of a single harmonic mode. Even the creation of real photons out of the vacuum, known as the dynamical Casimir effect, has been seen \cite{wilson:2011,hakonen:2012}.
 However,  LS modifications should naturally arise in electric circuits as the devices of interest can never be isolated from their surroundings, particularly from those with broadband spectral densities.~While this class of environments constitutes the most common one for solid-state systems, evidence for corresponding LS effects has proven elusive yet. Engineered environments such as those realized in atomic setups \cite{Myatt:2000, Kielpinski:2001} and studied for solid-state systems \cite{Gramich:2011, Solinas:2012, Murch:2012} may pose serious challenges in actual realizations though.

In this Letter, we propose a different scheme for the LS detection in a solid-state system weakly interacting with a broadband environment.  Instead of {\em engineering} the environment, the system is subject to a {\em strong and tunable driving field}. This drive can not be treated as a perturbation and the interaction between driven system and its environment is best described in terms of ``dressed states'' that recently have attracted much attention in superconducting circuits \cite{Oliver:2005,Wilson:2007, izmalkov:2008}. Under certain resonant conditions, the system-environment coupling is substantially enhanced by the presence of the drive, yielding a dynamic steady state which is largely determined by environmental features itself \cite{Gasparinetti:2013, stace:2005, stace:2013}. Within the same regime, the environment also induces a renormalization of the dressed-state energies (quasienergies). This renormalization defines the LS of the driven system. For atomic states in a resonant radiation field a LS has been discussed in \cite{jentschura:2003}, however, here we consider a generalized situation of  open quantum systems subject to arbitrary periodic driving. We find that the relative magnitude of the LS can exceed by far what is typically observed in static systems and exhibits specific scaling trends as a function of the drive parameters. These two features should make it easier to unambiguously identify the LS contribution.

While enhanced LSs should be observable in other driven solid-state systems as well \cite{Ramsay:2010, stace:2013,colless:2013}, here we focus on superconducting devices and consider circuits containing CPBs. This provides direct contact to recent experiments on driven CPBs used to realize a Mach-Zehnder interferometer \cite{Oliver:2005} and to operate as a charge pump \cite{niskanen:2003, Mottonen:2008, Gasparinetti:2011, Gasparinetti:2012}. Particularly, the latter situation reveals that the LS may also induce clear signatures in coherent charge currents.

\paragraph{LS of a driven two-level system.-} We start with a model, where a CPB subject to a general periodic drive is described by a driven two-level system (TLS) in terms of Pauli matrices, i.e.,
\begin{equation}
H_S(t)=-\frac{E}{2}\sigma_z + \vec{F}(t)\cdot \vec{\sigma}\, .
\label{eq:Rabihamiltonian}
\end{equation}
Here, $E$ is the level spacing of the bare system and $\vec{F}(t)=(F_x(t), F_y(t), F_z(t))$ is a driving field obeying $\vec{F}(t)=\vec{F}(t+2\pi/\Omega)$ with period $\Omega$.~This TLS interacts via $H_I= S\, \sum_j c_j (b_j^\dagger +b_j)$ with a reservoir of bosonic modes, i.e.\ $[b_j, b_j^{\dagger}]=1$, the distribution of which is characterized by a spectral function $J(\omega)=(\pi/\hbar)\sum_j c_j^2 \delta(\omega-\omega_j)$. For typical solid-state aggregates, this distribution has a broadband profile in contrast to single modes for high quality cavities. The system operator
\begin{equation}\label{eq:couplingS}
S(r)= \sin(r)\, \sigma_x +\cos(r)\, \sigma_z\,
\end{equation}
is chosen such as to capture both decoherence ($r=\pi/2$) or pure dephasing ($r=0$).

A powerful approach to treat periodically driven quantum dynamics is given by the Floquet formalism \cite{Grifoni:1998}.~One starts from a complete set of solutions of the time-dependent Schr\"odinger equation for $H_S(t)=H_S(t+ 2\pi/\Omega)$ given by the Floquet states $|\Psi_\alpha(t)\rangle= {\rm e}^{-i\epsilon_\alpha t/\hbar}|\Phi_\alpha(t)\rangle$, where the Floquet modes $|\Phi_\alpha\rangle$ satisfy
$|\Phi_\alpha(t)\rangle=|\Phi_\alpha(t+2\pi/\Omega)\rangle$.~The quasienergies $\epsilon_\alpha$ play the role of dressed state energies and are only defined mod$[\hbar\Omega]$. The Floquet description manifestly takes into account the fundamental and all higher harmonics and thus also applies to {\em arbitrary strong driving far from resonance}.

In case of weak dissipation, the Floquet formalism can be consistently combined with second order perturbation theory to arrive at a Born-Markov-type master equation for the reduced dynamics of the system  $\rho(t)$ \cite{Grifoni:1998}.~After performing a partial secular approximation (PSA) \cite{Gasparinetti:2013}, this master equation becomes time-independent in the basis of the Floquet modes and takes in the Schr\"odinger picture the form
\begin{equation}
\dot{\rho}_{\alpha\beta}(t) = -i (\omega_{\alpha\beta}-\delta\omega_{\alpha\beta})\,\rho_{\alpha\beta}(t)+ \sum_{\gamma,\delta}\mathcal{R}_{\alpha\beta\gamma\delta}\,\rho_{\gamma\delta}(t)
\label{eq:mastereq}
\end{equation}
with transition frequencies $\omega_{\alpha \beta}=(\epsilon_\alpha-\epsilon_\beta)/\hbar$. The Redfield tensor $\mathcal{R}_{\alpha\beta\gamma\delta}$ captures decoherence as well as dephasing and formally couples equations for the populations (diagonal elements of the density) and the coherences (off-diagonal ones) (see Supplemental Material \cite{epaps}).
In extension to previous treatments \cite{Grifoni:1998, Russomanno:2011, Gasparinetti:2013}, Eq.~(\ref{eq:mastereq}) contains the reservoir induced renormalization $\delta\omega_{\alpha\beta}$ of transition frequencies.
At sufficiently low temperatures, it is dominated by environmental zero-point fluctuations and then constitutes the LS for a driven quantum system. This is the main focus of this work which, to our knowledge and despite of its relevance for ongoing experiments, has not been addressed yet.

Specifically, in case of a TLS at zero temperature we find (see \cite{epaps})
\begin{equation}\label{eq:Lambshift}
\delta \omega_{12}=\frac{1}{\pi \hbar}\sum_{k=-\infty}^\infty |X_{21,k}^{(r)}|^2  \,[G(\Delta_{21,k})-G(-\Delta_{21,k})]
\end{equation}
which contains dressed transition frequencies $\Delta_{\alpha\beta, k} = (\epsilon_{\alpha}-\epsilon_{\beta})/\hbar + k \Omega$ and coupling matrix elements
\[
X_{\alpha\beta, k}^{(r)}= \frac{\Omega}{2\pi}\int_{0}^{2\pi/\Omega} dt\, {\rm e}^{-ik\Omega t}\langle \Phi_{\alpha}(t)|S(r)|\Phi_{\beta}(t)\rangle\, .
\]
 Reservoir properties are encoded in the principal value integral $G(z)=-\mathcal{P}\int_0^\infty d\omega\, J(\omega)/(\omega-z)$.~These results apply to any periodic driving, spectral function and coupling operator (\ref{eq:couplingS}). Thermal corrections to (\ref{eq:mastereq}) and to (\ref{eq:Lambshift}), also known as ac-Stark shift, play a role if $\hbar\Delta_{\alpha \beta, k}$ is comparable to the thermal energy. As we will discuss in more detail below, this imposes conditions on driving frequencies and amplitudes to sufficiently exceed $1/\beta=k_{\rm B} T$.

The master equation (\ref{eq:mastereq}) substantially differs from that for undriven systems in that the effective system-bath coupling is given by $\eta \Omega/|\omega_{12}|$ \cite{Gasparinetti:2013} with a bare system-bath coupling parameter $\eta$. Due to the tunability of $\omega_{12}$ via driving amplitude or frequency, environmental induced terms can thus even dominate the dynamics without deteriorating the validity of (\ref{eq:mastereq}) \cite{Gasparinetti:2013}.

\begin{figure}[h]
\includegraphics[width=1.02\columnwidth]{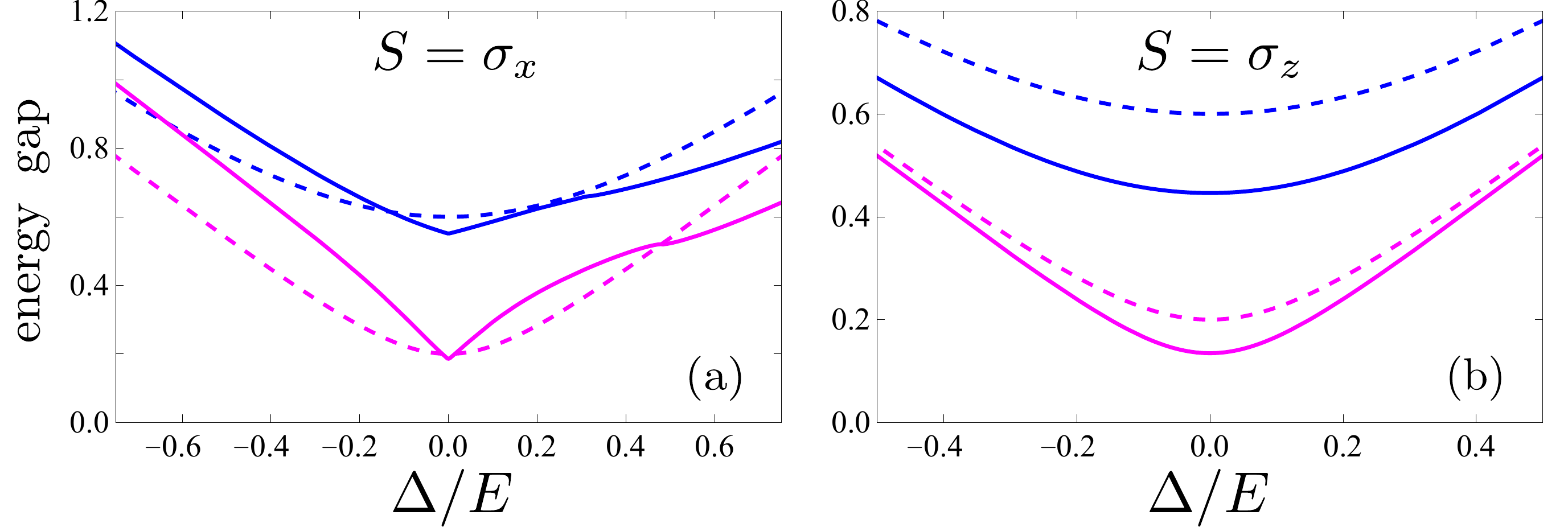}
\caption{Quasienergy gap $\hbar\omega_{12}/E$ with (solid) and without (dashed) LS according to (\ref{lamb}) as a function of detuning $\Delta$ for driving amplitudes $|A|/E=0.1\,{\rm (magenta)\, and}\, 0.3\, {\rm (blue)}$, and different coupling mechanisms to the bath: (a) $S=\sigma_x$  and (b) $S=\sigma_z$. Damping parameters are $\eta=0.1$ and cut-off frequency $\hbar\omega_c/E=60$.}
\label{fig:rengap}
\end{figure}

\paragraph{Strongly driven CPB in an Ohmic reservoir.-}To illustrate the above findings and to discuss their relevance in actual experiments, we now consider a specific model. The bath is assumed to be Ohmic-like, i.e., $J(\omega)=\eta \hbar \omega \exp(-\omega/\omega_c)$ with coupling constant $\eta$ and a large cut-off frequency $\omega_c$, and the external drive in (\ref{eq:Rabihamiltonian}) is taken as $\vec{F}(t)=A (\cos(\Omega t), -\sin(\Omega t), 0)$ with drive amplitude $A$. The Floquet states can then explicitly be calculated (see \cite{epaps}) with quasienergies $\epsilon_{1,2}=(\Delta\pm\hbar\omega_R)/2$, where $\Delta=E-\hbar\Omega$ is the detuning and $\omega_R=\sqrt{\Delta^2 + 4|A|^2}/\hbar$ the Rabi frequency. In typical realizations $\Omega\gg \eta\omega_R$ which justifies the PSA in (\ref{eq:mastereq}). This model is also known as the semiclassical Rabi model, first used to describe optical transitions of atoms \cite{scully:1997}.

 Transparent expressions for the LS (\ref{eq:Lambshift}) with the coupling $S(r)$ in (\ref{eq:couplingS}) are  obtained in the two limiting cases with mixing angles $r=\pi/2$ (transversal coupling inducing decoherence) and $r=0$ (longitudinal coupling inducing dephasing).  One gains $\delta \omega_{12}=(\eta \omega_c/\pi)\, \Lambda^{(r)}$, where
\begin{eqnarray}\label{lamb}
\Lambda^{(0)}&=&- g(\omega_R/\omega_c) \sin^2(2\theta)\nonumber\\
\Lambda^{(\pi/2)}&=& g(\omega_-/\omega_c) \sin^4(\theta)+g(\omega_+/\omega_c) \cos^4(\theta)\,
\end{eqnarray}
with $\omega_\pm=\pm \Omega-\omega_R$ and $\tan(2\theta)= 2 |A|/\Delta$. Reservoir zero-point fluctuations enter through $g(x)= x\,[ {\rm Ei}(x)\,{\rm e}^{-x}+{\rm Ei}(-x)\,{\rm e}^{x}]$, the asymptotics of which $g(|x|\ll1)\approx 2x\, {\rm ln}(|x|)$ reflects the logarithmic behavior known from the atomic LS \cite{Lamb:1947}. Apparently, the LS (\ref{lamb}) also carries information about the system (CPB)-reservoir coupling mechanism.~It turns out that for $\hbar\omega_c/E\gg 1$, results discussed below depend on the cut-off only very weakly.

Upon tuning the drive amplitude and/or frequency, the LS thus displays a qualitatively different behavior in contrast to the bare quasienergy gap (Fig.~\ref{fig:rengap}). Namely,  for transversal coupling ($S=\sigma_x$), we find for the screened transition frequency a pronounced asymmetry with respect to negative and positive detuning (Fig.~\ref{fig:rengap}a).~A measurement of this deviation from the linear scaling would be a clear signature of the presence of environmental ground state fluctuations. On the contrary, longitudinal coupling ($S=\sigma_z$) leads to a smoother structure and less deviations from the bare gap, see Fig.~\ref{fig:rengap}b.

\begin{figure}[h]
\includegraphics[width=1.02\columnwidth]{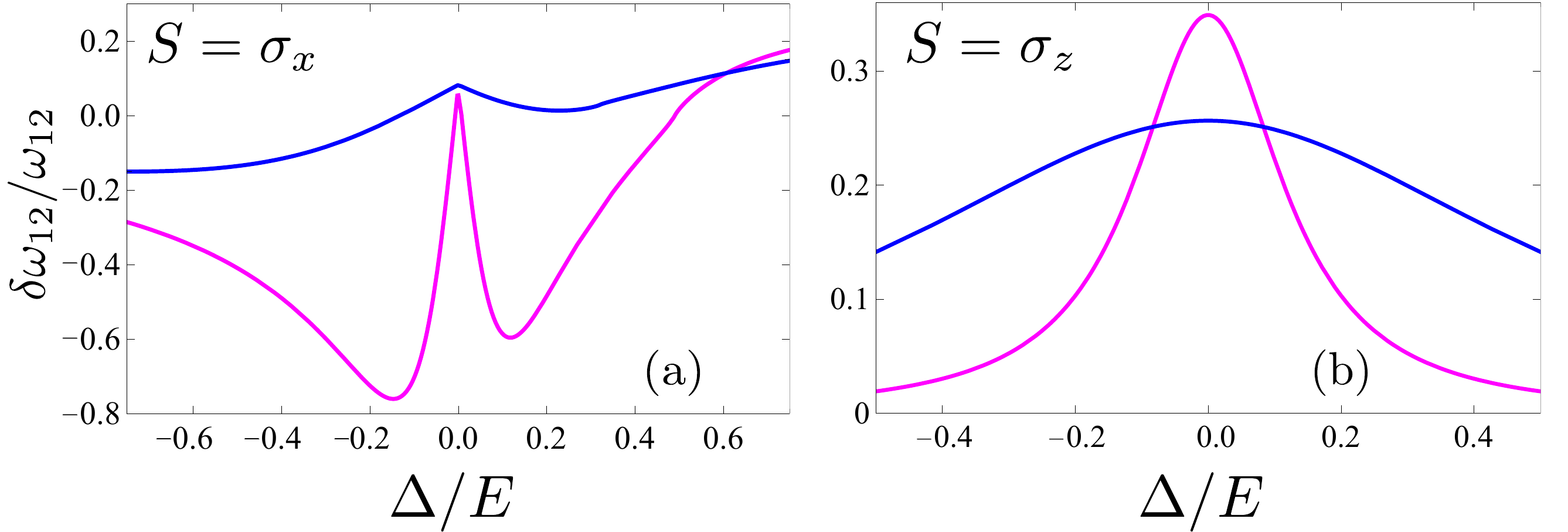}
\caption{Relative magnitude of the LS $\delta\omega_{12}/\omega_{12}$ for (a) transversal ($\sigma_x$) and (b) longitudinal ($\sigma_z$) coupling vs.\ detuning $\Delta$ and for different values of the driving amplitude $|A|/E$: 0.07 (magenta), 0.3 (blue).  Other parameters are as in Fig.~\ref{fig:rengap}.}
\label{fig:LSscaledwithquasiens}
\end{figure}
To have a more quantitative estimate of the impact of the LS and, in particular, to reveal optimal parameter regimes for detection with respect to signal strength and suppression of thermal fluctuations, in Fig.~\ref{fig:LSscaledwithquasiens}, we plot the relative magnitude of the LS compared to the bare transition frequency when the detuning is varied. In fact, the external driving increases the ratio $|\delta \omega_{12} / \omega_{12}|$ for both coupling schemes far above the usual ratio of a few percent \cite{Fragner:2008}. The dressed system exchanges energy quanta with the bath with an effective coupling $\eta \Omega/\omega_R$ which due to $\Omega\gg\omega_R$ can be strong compared to the static situation. We mention that this enhancement also applies to the limiting situation of a single mode reservoir (cavity), see \cite{epaps}.

Experimentally, strong signals for the LS in both cases of transversal and longitudinal noise can be expected for weak to moderate detuning $|\Delta|/E \lesssim 0.3$ (Fig.~\ref{fig:LSscaledwithquasiens}).~With respect to low thermal fluctuations,
the conditions are different. For $\sigma_x$-coupling, Fig.~\ref{fig:LSscaledwithquasiens}a, thermal noise is suppressed if $\Omega\hbar\beta\gg 1$ (see \cite{epaps}) which is easily fulfilled also for relatively weak driving $|A|/E\ll 1$, where the LS is pronounced. Instead, for $\sigma_z$-coupling, Fig.~\ref{fig:LSscaledwithquasiens}b, due to the constraint $\omega_R\hbar\beta\gg 1$ \cite{epaps} stronger driving is required which, as shown, supports enhanced LS yet.

\begin{figure}[h]
\includegraphics[width=1.02\columnwidth]{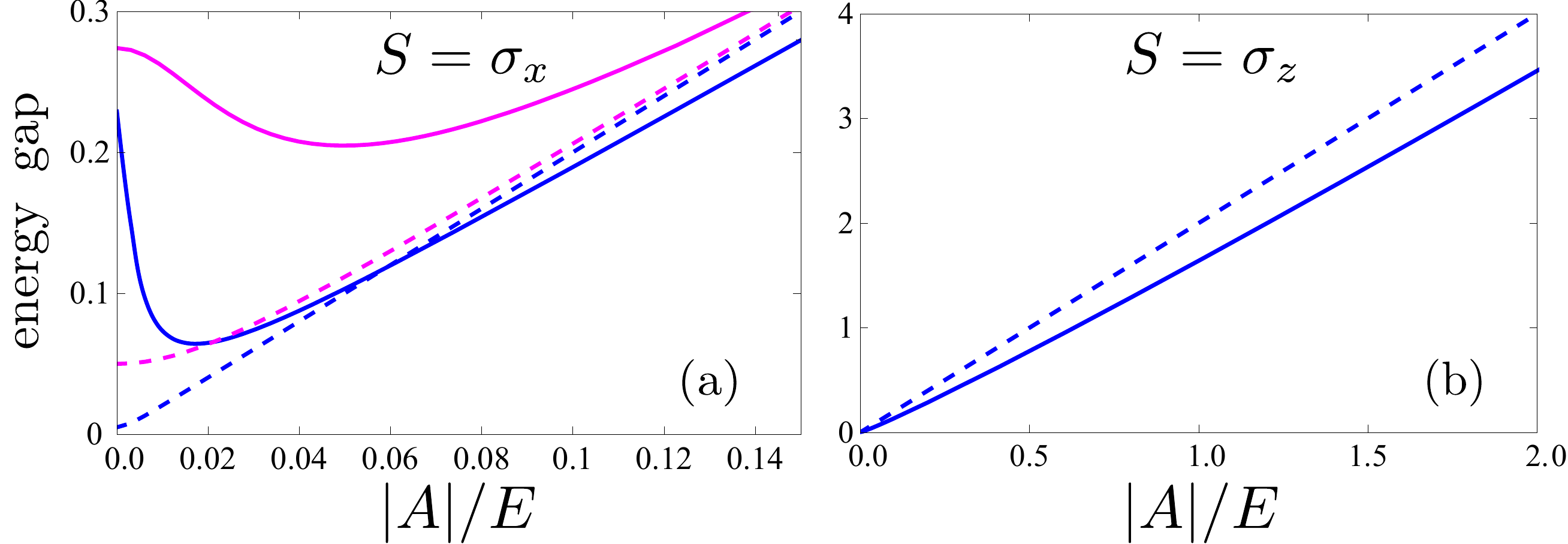}
\caption{Same as in Fig.~\ref{fig:rengap}, but vs.\ driving amplitude $|A|$ close to the resonance with $|\Delta|/E$: 0.005 (blue) and 0.05 (magenta). The corresponding curves in (b) cannot be resolved on this scale.}
\label{fig:driveamplitude}
\end{figure}
An alternative way to measure an enhanced LS by varying the drive amplitude $A$ is presented in Fig.~\ref{fig:driveamplitude}: For longitudinal coupling (b) one finds a smooth behavior, where close to resonance the dependence on $|\Delta|$ disappears and $\hbar\delta\omega_{12}\sim |A|$. The situation for transversal coupling (a) is quite different. In this case, according to $\Lambda^{(\pi/2)}$ in (\ref{lamb}) one has  $\delta\omega_{12}(|A|/|\Delta|\ll 1)\approx (2\Omega \eta/\pi)\,{\rm ln}(\Omega/\omega_c)$, while then the bare gap $\hbar\omega_{12}\approx |\Delta|$. This leads to a highly non-monotonic dependence of the full gap on the driving amplitude and allows to distinguish between the prevalence of either of the coupling mechanisms.~The discontinuity of the LS in the limits $|\Delta|=0, |A|/E>0$ (seen in Fig.~\ref{fig:LSscaledwithquasiens}a) and $|\Delta|/E>0, A=0$ (seen in Fig.~\ref{fig:driveamplitude}a) is due to the $\theta$-dependence of $\Lambda^{(\pi/2)}$.

\paragraph{LS detection.-} We now discuss how in recent experiments with driven CPBs the model (\ref{eq:Rabihamiltonian}) together with various system-bath couplings (\ref{eq:couplingS}) is realized and how the LS could be retrieved. In \cite{Oliver:2005,Wilson:2007, wilson:2010}, a CPB with tunable charging energy $E_{C}$ and Josephson energy $E_J$ is subject to a microwave field and embedded in an environment which dominantly induces charge fluctuations.~The Hamiltonian takes for large photon fields the form $H_{S,\rm CPB}(t)=-\frac{1}{2} E_{C}\tau_z -\frac{1}{2} E_J \tau_x - \lambda \cos(\omega t) \tau_z$ with Pauli matrices $\{\tau_i\}$ and $\tau_z$-coupling to the bath.
In the general situation, the eigenstate representation of the CPB leads to (\ref{eq:Rabihamiltonian}) with $E=\sqrt{E_C^2+E_J^2}$ and with both the system-bath coupling and the drive fixed by the angle $r=\arctan(E_J/E_C)$ in (\ref{eq:couplingS}) where $\vec{F}(t)=\lambda \cos(\omega t) (\sin(r), 0,-\cos(r))$. Simplifications are achieved in two limiting cases.~(i) At charge degeneracy $E_C=0$, one has $E\approx E_J$ and $\vec{F}(t)=\lambda \cos(\omega t)(1,0,0)$ together with $S(\pi/2)$ (transversal coupling). For $|\lambda|\ll \hbar\omega\sim E_J$, within the PSA, the drive even reduces to the model in (\ref{lamb}).  (ii) For $E_C\neq 0$ but strong driving $ \hbar\omega\gg \lambda\gg E_J$, effectively, longitudinal noise ($r=0$) is obtained. Then, instead of working in the CPB eigenbasis, one applies a dressed tunneling picture \cite{Oliver:2005,epaps}. Close to the $n$-th order photon resonance $E_C=n\hbar\omega$, the PSA consistently reduces $H_{S, \rm CPB}(t)$ to the $n$-photon sector and one arrives at (\ref{lamb}) with $E=E_C$,  $A=-(E_J/2)\, J_n(2\lambda/\hbar\omega), \Omega=n\omega$, and $r=0$ (see \cite{epaps}).

Now, after preparation of the CPB, its steady state in presence of the microwave drive [$\dot{\rho}_{\alpha\beta}=0$  in (\ref{eq:mastereq})] is probed with a weak pulse $H_P=\mu_P \cos(\omega_P t)\, \sigma_z, \mu_P\ll A$, to access the resonance at $\omega_{12}-\delta \omega_{12}=\omega_R-(\eta\omega_c/\pi)\Lambda^{(r)}$. The LS then appears in the absorption spectrum upon varying either the detuning (Fig.~\ref{fig:rengap}) or the amplitude (Fig.~\ref{fig:driveamplitude}) of the pump field (cf.~\cite{Silveri:2013}). For typical experimental parameters from \cite{wilson:2010} $E/\hbar\approx 10$\,GHz and $\eta\approx 0.05, \hbar\omega_c/E\approx 60$, $E\beta\approx 7$, one finds for $\sigma_x$-coupling $|\delta\omega_{12}(\hbar\Omega/E=1.1, |A|/E=0.1)|/\omega_R\approx 0.2$ with $\Omega\hbar\beta\approx 8$, while for $\sigma_z$-coupling $|\delta\omega_{12}(\hbar\Omega/E=1.3, |A|/E=0.3)|/\omega_R\approx 0.1$ with still sufficiently weak thermal noise $\omega_R\hbar\beta\approx 5$. Already these LS values are at least one order of magnitude larger than in the static case \cite{Fragner:2008} and can further be enhanced by optimized circuit designs and detection protocols.

\begin{figure}[h]
\includegraphics[width=1.0\columnwidth]{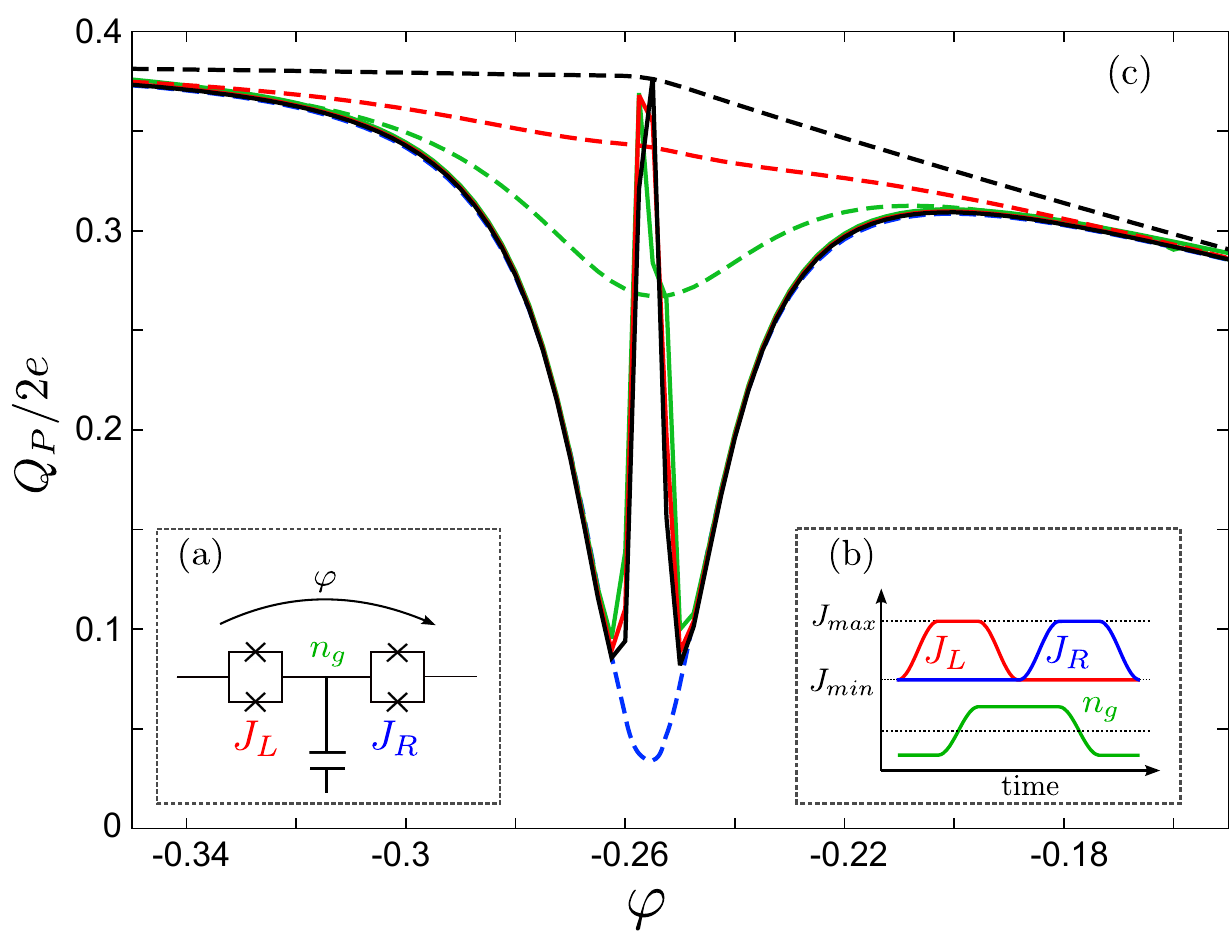}
\caption{(a) Schematic circuit diagram for the `sluice' and (b) driving protocol with the three time-dependent control parameters $J_{L,R}$ and $n_g$ for one pumping cycle. (c): Pumped charge $Q_P$ with (solid) and without (dashed) LS contribution vs. phase $\varphi$ for different values of the system-bath coupling: $\eta=0.001$ (blue), $\eta=0.005$ (green), $\eta=0.01$ (red) and $\eta=0.05$ (black). Parameters are chosen according to \cite{niskanen:2003}: drive time $\tau=$ 1\,ns, $E_C= $1\,K, $\omega_c=$ 100\,GHz, $\Delta n_g=n_g-0.5=$ 0.2, $J_{\rm max}= 0.1\,E_C$, $J_{\rm min}= 10^{-3}\,J_{\rm max}$.}
\label{fig:QPzoom}
\end{figure}

\paragraph{Cooper pair pump.-}
The devices discussed up to this point are tailored to externally manipulate their level structure. However, the LS may have also profound impact in superconducting circuits where transport properties are addressed. A specific example is a charge pump in form of the Cooper pair sluice \cite{niskanen:2003, Niskanen:2005} sketched in Fig.~\ref{fig:QPzoom}a.
 It consists of a single superconducting island (CPB) separated by two SQUIDs with tunable Josephson energies $J_{L,R}(t)$. A third control parameter is provided by the gate charge $n_g(t)$ capacitively coupled to the island. Full control of the quantum system is guaranteed via the three experimentally accessible parameters $J_{L,R}$ and $n_g$ which allow for charge pumping when steered in a periodic protocol (see Fig.~\ref{fig:QPzoom}b).
In the charging regime $E_C \gg {\rm max}\{J_L, J_R\}$ and close to a half integer of the gate charge $n_g(t)$, this device is described by a pseudospin-Hamiltonian \cite{niskanen:2003, Solinas:2010}
\begin{equation}
H_S(t)=-\frac{1}{2}\,\vec{B}(t)\cdot \vec{\sigma}\, ,
\end{equation}
where  $B_x(t)=J_+(t)\cos(\frac{\varphi}{2})$, $B_y(t)=J_-(t)\sin(\frac{\varphi}{2})$, $B_z(t)=E_C[2n_g(t)-1]$ and $J_{\pm}(t)=J_L(t)\pm J_R(t)$.  The total superconducting phase difference across the sluice is denoted by $\varphi$.
Dominant noise sources are charge fluctuations implying a $\sigma_z$-coupling to environmental modes.

This system has a more complex structure than the Rabi model as it includes transversal and longitudinal driving with many higher harmonics. Its main observable is the charge $Q_P$  pumped through the sluice during sequences of driving cycles.
Numerical results in steady state based on (\ref{eq:mastereq}) are depicted in Fig.~\ref{fig:QPzoom}c with and without the LS versus the phase $\varphi$ across the sluice. The latter one can be adjusted by an external magnetic field.

The pumped charge displays a very sensitive dependence on the phase difference which in turn determines the energy splitting.~Close to a degeneracy of the quasienergies at $\varphi_c\approx -0.26$, one enters a regime where environmental effects on $Q_P$ are strong and of order $\eta\Omega/|\omega_{12}-\delta\omega_{12}|$ \cite{Gasparinetti:2013}.
In contrast to the bare situation, however, the pumped charge including the LS depends only very weakly on the coupling parameter $\eta$. Namely, away from degeneracy (away from the peak), the LS enhances the level splitting so that renormalized dressed system properties prevail against decoherence. Noise induced transitions between the energy levels are thus suppressed and the pumped charge follows the bare one for $\eta=0.001$. Within the domain of the bare crossing
 $\omega_{12} \approx 0$ (range of the peak), the steady state of  (\ref{eq:mastereq}) is completely determined by reservoir quantities, i.e., the LS and the Redfield tensor. Since both are proportional to $\eta$, the friction parameter drops out of the steady state equation. Apparently, within this latter range the predictions for $Q_P$ with LS qualitatively deviate from those without LS giving rise to a peak instead of a dip. This verifies the pronounced impact of vacuum fluctuations also on transport properties of superconducting circuits.

\paragraph{Conclusion.-} We have analyzed the impact of environmental zero-point fluctuations with broadband spectral densities in strongly driven quantum systems. Depending on the drive parameters, the relative strength of the system-reservoir coupling is enhanced and, thus, the induced LS increased to an extent unreachable in standard experiments. This LS displays distinctive signatures as a function of the driving amplitude and/or frequency.
The predicted effect should be accessible in many solid-state systems, particularly in current superconducting devices.
 Specific detection schemes have been discussed for circuits with driven artificial atoms and controlled Cooper pair charge flow. The proposed protocols would shed new light on the impact of broadband environments on quantum systems at cryogenic temperatures which is completely absent in the classical regime.
\paragraph{Acknowledgements.-} The authors would like to thank M. G\"unther and S. Pugnetti for fruitful discussions.
This work has been supported by the German Science Foundation (DFG) within SFB/TRR-21 and AN336/6 as well as by the European Community's Seventh Framework Programme (FP7/2007-2013) under grant agreement No. 228464 (MICROKELVIN). We gratefully acknowledge also financial support from the DAAD (V.~G.) and the Finnish Graduate School in Nanoscience (S.~G.). P.~S. thanks for the support from FIRB - Futuro in Ricerca 2013 under Grant No. RBFR1379UX and FIRB 2012 under Grant No. RBFR1236VV HybridNanoDev.

\bibliographystyle{apsrev4-1}
%\bibliography{bibLamb}

\begin{thebibliography}{40}%
\makeatletter
\providecommand \@ifxundefined [1]{%
 \@ifx{#1\undefined}
}%
\providecommand \@ifnum [1]{%
 \ifnum #1\expandafter \@firstoftwo
 \else \expandafter \@secondoftwo
 \fi
}%
\providecommand \@ifx [1]{%
 \ifx #1\expandafter \@firstoftwo
 \else \expandafter \@secondoftwo
 \fi
}%
\providecommand \natexlab [1]{#1}%
\providecommand \enquote  [1]{``#1''}%
\providecommand \bibnamefont  [1]{#1}%
\providecommand \bibfnamefont [1]{#1}%
\providecommand \citenamefont [1]{#1}%
\providecommand \href@noop [0]{\@secondoftwo}%
\providecommand \href [0]{\begingroup \@sanitize@url \@href}%
\providecommand \@href[1]{\@@startlink{#1}\@@href}%
\providecommand \@@href[1]{\endgroup#1\@@endlink}%
\providecommand \@sanitize@url [0]{\catcode `\\12\catcode `\$12\catcode
  `\&12\catcode `\#12\catcode `\^12\catcode `\_12\catcode `\%12\relax}%
\providecommand \@@startlink[1]{}%
\providecommand \@@endlink[0]{}%
\providecommand \url  [0]{\begingroup\@sanitize@url \@url }%
\providecommand \@url [1]{\endgroup\@href {#1}{\urlprefix }}%
\providecommand \urlprefix  [0]{URL }%
\providecommand \Eprint [0]{\href }%
\providecommand \doibase [0]{http://dx.doi.org/}%
\providecommand \selectlanguage [0]{\@gobble}%
\providecommand \bibinfo  [0]{\@secondoftwo}%
\providecommand \bibfield  [0]{\@secondoftwo}%
\providecommand \translation [1]{[#1]}%
\providecommand \BibitemOpen [0]{}%
\providecommand \bibitemStop [0]{}%
\providecommand \bibitemNoStop [0]{.\EOS\space}%
\providecommand \EOS [0]{\spacefactor3000\relax}%
\providecommand \BibitemShut  [1]{\csname bibitem#1\endcsname}%
\let\auto@bib@innerbib\@empty
%</preamble>
\bibitem [{\citenamefont {Lamb}\ and\ \citenamefont
  {Retherford}(1947)}]{Lamb:1947}%
  \BibitemOpen
  \bibfield  {author} {\bibinfo {author} {\bibfnamefont {W.~E.}\ \bibnamefont
  {Lamb}}\ and\ \bibinfo {author} {\bibfnamefont {R.~C.}\ \bibnamefont
  {Retherford}},\ }\href {\doibase 10.1103/PhysRev.72.241} {\bibfield
  {journal} {\bibinfo  {journal} {Phys. Rev.}\ }\textbf {\bibinfo {volume}
  {72}},\ \bibinfo {pages} {241} (\bibinfo {year} {1947})}\BibitemShut
  {NoStop}%
\bibitem [{\citenamefont {Heinzen}\ and\ \citenamefont
  {Feld}(1987)}]{Heinzen:1987}%
  \BibitemOpen
  \bibfield  {author} {\bibinfo {author} {\bibfnamefont {D.~J.}\ \bibnamefont
  {Heinzen}}\ and\ \bibinfo {author} {\bibfnamefont {M.~S.}\ \bibnamefont
  {Feld}},\ }\href {\doibase 10.1103/PhysRevLett.59.2623} {\bibfield  {journal}
  {\bibinfo  {journal} {Phys. Rev. Lett.}\ }\textbf {\bibinfo {volume} {59}},\
  \bibinfo {pages} {2623} (\bibinfo {year} {1987})}\BibitemShut {NoStop}%
\bibitem [{\citenamefont {Brune}\ \emph {et~al.}(1994)\citenamefont {Brune},
  \citenamefont {Nussenzveig}, \citenamefont {Schmidt-Kaler}, \citenamefont
  {Bernardot}, \citenamefont {Maali}, \citenamefont {Raimond},\ and\
  \citenamefont {Haroche}}]{Brune:1994}%
  \BibitemOpen
  \bibfield  {author} {\bibinfo {author} {\bibfnamefont {M.}~\bibnamefont
  {Brune}}, \bibinfo {author} {\bibfnamefont {P.}~\bibnamefont {Nussenzveig}},
  \bibinfo {author} {\bibfnamefont {F.}~\bibnamefont {Schmidt-Kaler}}, \bibinfo
  {author} {\bibfnamefont {F.}~\bibnamefont {Bernardot}}, \bibinfo {author}
  {\bibfnamefont {A.}~\bibnamefont {Maali}}, \bibinfo {author} {\bibfnamefont
  {J.~M.}\ \bibnamefont {Raimond}}, \ and\ \bibinfo {author} {\bibfnamefont
  {S.}~\bibnamefont {Haroche}},\ }\href {\doibase 10.1103/PhysRevLett.72.3339}
  {\bibfield  {journal} {\bibinfo  {journal} {Phys. Rev. Lett.}\ }\textbf
  {\bibinfo {volume} {72}},\ \bibinfo {pages} {3339} (\bibinfo {year}
  {1994})}\BibitemShut {NoStop}%
\bibitem [{\citenamefont {Marrocco}\ \emph {et~al.}(1998)\citenamefont
  {Marrocco}, \citenamefont {Weidinger}, \citenamefont {Sang},\ and\
  \citenamefont {Walther}}]{Marrocco:1998}%
  \BibitemOpen
  \bibfield  {author} {\bibinfo {author} {\bibfnamefont {M.}~\bibnamefont
  {Marrocco}}, \bibinfo {author} {\bibfnamefont {M.}~\bibnamefont {Weidinger}},
  \bibinfo {author} {\bibfnamefont {R.~T.}\ \bibnamefont {Sang}}, \ and\
  \bibinfo {author} {\bibfnamefont {H.}~\bibnamefont {Walther}},\ }\href
  {\doibase 10.1103/PhysRevLett.81.5784} {\bibfield  {journal} {\bibinfo
  {journal} {Phys. Rev. Lett.}\ }\textbf {\bibinfo {volume} {81}},\ \bibinfo
  {pages} {5784} (\bibinfo {year} {1998})}\BibitemShut {NoStop}%
\bibitem [{\citenamefont {Haroche}\ and\ \citenamefont
  {Raimond}(2006)}]{Haroche:2006}%
  \BibitemOpen
  \bibfield  {author} {\bibinfo {author} {\bibfnamefont {S.}~\bibnamefont
  {Haroche}}\ and\ \bibinfo {author} {\bibfnamefont {J.-M.}\ \bibnamefont
  {Raimond}},\ }\href@noop {} {\emph {\bibinfo {title} {{Exploring the Quantum:
  Atoms, Cavities, and Photons}}}}\ (\bibinfo  {publisher} {Oxford University
  Press, New York},\ \bibinfo {year} {2006})\BibitemShut {NoStop}%
\bibitem [{\citenamefont {Wallraff}\ \emph {et~al.}(2004)\citenamefont
  {Wallraff}, \citenamefont {Schuster}, \citenamefont {Blais}, \citenamefont
  {Frunzio}, \citenamefont {Huang}, \citenamefont {Majer}, \citenamefont
  {Kumar}, \citenamefont {Girvin},\ and\ \citenamefont
  {Schoelkopf}}]{Wallraff:2004}%
  \BibitemOpen
  \bibfield  {author} {\bibinfo {author} {\bibfnamefont {A.}~\bibnamefont
  {Wallraff}}, \bibinfo {author} {\bibfnamefont {D.~I.}\ \bibnamefont
  {Schuster}}, \bibinfo {author} {\bibfnamefont {A.}~\bibnamefont {Blais}},
  \bibinfo {author} {\bibfnamefont {L.}~\bibnamefont {Frunzio}}, \bibinfo
  {author} {\bibfnamefont {R.-S.}\ \bibnamefont {Huang}}, \bibinfo {author}
  {\bibfnamefont {J.}~\bibnamefont {Majer}}, \bibinfo {author} {\bibfnamefont
  {S.}~\bibnamefont {Kumar}}, \bibinfo {author} {\bibfnamefont {S.~M.}\
  \bibnamefont {Girvin}}, \ and\ \bibinfo {author} {\bibfnamefont {R.~J.}\
  \bibnamefont {Schoelkopf}},\ }\href@noop {} {\bibfield  {journal} {\bibinfo
  {journal} {Nature}\ }\textbf {\bibinfo {volume} {431}},\ \bibinfo {pages}
  {162} (\bibinfo {year} {2004})}\BibitemShut {NoStop}%
\bibitem [{\citenamefont {Schuster}\ \emph {et~al.}(2007)\citenamefont
  {Schuster}, \citenamefont {Houck}, \citenamefont {Schreier}, \citenamefont
  {Wallraff}, \citenamefont {Gambetta}, \citenamefont {Blais}, \citenamefont
  {Frunzio}, \citenamefont {Majer}, \citenamefont {Johnson}, \citenamefont
  {Devoret}, \citenamefont {Girvin},\ and\ \citenamefont
  {Schoelkopf}}]{Schuster:2007}%
  \BibitemOpen
  \bibfield  {author} {\bibinfo {author} {\bibfnamefont {D.~I.}\ \bibnamefont
  {Schuster}}, \bibinfo {author} {\bibfnamefont {A.~A.}\ \bibnamefont {Houck}},
  \bibinfo {author} {\bibfnamefont {J.~A.}\ \bibnamefont {Schreier}}, \bibinfo
  {author} {\bibfnamefont {A.}~\bibnamefont {Wallraff}}, \bibinfo {author}
  {\bibfnamefont {J.~M.}\ \bibnamefont {Gambetta}}, \bibinfo {author}
  {\bibfnamefont {A.}~\bibnamefont {Blais}}, \bibinfo {author} {\bibfnamefont
  {L.}~\bibnamefont {Frunzio}}, \bibinfo {author} {\bibfnamefont
  {J.}~\bibnamefont {Majer}}, \bibinfo {author} {\bibfnamefont
  {B.}~\bibnamefont {Johnson}}, \bibinfo {author} {\bibfnamefont {M.~H.}\
  \bibnamefont {Devoret}}, \bibinfo {author} {\bibfnamefont {S.~M.}\
  \bibnamefont {Girvin}}, \ and\ \bibinfo {author} {\bibfnamefont {R.~J.}\
  \bibnamefont {Schoelkopf}},\ }\href@noop {} {\bibfield  {journal} {\bibinfo
  {journal} {Nature}\ }\textbf {\bibinfo {volume} {445}},\ \bibinfo {pages}
  {515} (\bibinfo {year} {2007})}\BibitemShut {NoStop}%
\bibitem [{\citenamefont {Blais}\ \emph {et~al.}(2007)\citenamefont {Blais},
  \citenamefont {Gambetta}, \citenamefont {Wallraff}, \citenamefont {Schuster},
  \citenamefont {Girvin}, \citenamefont {Devoret},\ and\ \citenamefont
  {Schoelkopf}}]{Blais:2007}%
  \BibitemOpen
  \bibfield  {author} {\bibinfo {author} {\bibfnamefont {A.}~\bibnamefont
  {Blais}}, \bibinfo {author} {\bibfnamefont {J.}~\bibnamefont {Gambetta}},
  \bibinfo {author} {\bibfnamefont {A.}~\bibnamefont {Wallraff}}, \bibinfo
  {author} {\bibfnamefont {D.~I.}\ \bibnamefont {Schuster}}, \bibinfo {author}
  {\bibfnamefont {S.~M.}\ \bibnamefont {Girvin}}, \bibinfo {author}
  {\bibfnamefont {M.~H.}\ \bibnamefont {Devoret}}, \ and\ \bibinfo {author}
  {\bibfnamefont {R.~J.}\ \bibnamefont {Schoelkopf}},\ }\href {\doibase
  10.1103/PhysRevA.75.032329} {\bibfield  {journal} {\bibinfo  {journal} {Phys.
  Rev. A}\ }\textbf {\bibinfo {volume} {75}},\ \bibinfo {pages} {032329}
  (\bibinfo {year} {2007})}\BibitemShut {NoStop}%
\bibitem [{\citenamefont {Schuster}\ \emph {et~al.}(2005)\citenamefont
  {Schuster}, \citenamefont {Wallraff}, \citenamefont {Blais}, \citenamefont
  {Frunzio}, \citenamefont {Huang}, \citenamefont {Majer}, \citenamefont
  {Girvin},\ and\ \citenamefont {Schoelkopf}}]{Schuster:2005}%
  \BibitemOpen
  \bibfield  {author} {\bibinfo {author} {\bibfnamefont {D.~I.}\ \bibnamefont
  {Schuster}}, \bibinfo {author} {\bibfnamefont {A.}~\bibnamefont {Wallraff}},
  \bibinfo {author} {\bibfnamefont {A.}~\bibnamefont {Blais}}, \bibinfo
  {author} {\bibfnamefont {L.}~\bibnamefont {Frunzio}}, \bibinfo {author}
  {\bibfnamefont {R.-S.}\ \bibnamefont {Huang}}, \bibinfo {author}
  {\bibfnamefont {J.}~\bibnamefont {Majer}}, \bibinfo {author} {\bibfnamefont
  {S.~M.}\ \bibnamefont {Girvin}}, \ and\ \bibinfo {author} {\bibfnamefont
  {R.~J.}\ \bibnamefont {Schoelkopf}},\ }\href {\doibase
  10.1103/PhysRevLett.94.123602} {\bibfield  {journal} {\bibinfo  {journal}
  {Phys. Rev. Lett.}\ }\textbf {\bibinfo {volume} {94}},\ \bibinfo {pages}
  {123602} (\bibinfo {year} {2005})}\BibitemShut {NoStop}%
\bibitem [{\citenamefont {Fink}\ \emph {et~al.}(2008)\citenamefont {Fink},
  \citenamefont {G\"oppl}, \citenamefont {Baur}, \citenamefont {Bianchetti},
  \citenamefont {Leek}, \citenamefont {Blais},\ and\ \citenamefont
  {Wallraff}}]{Fink:2008}%
  \BibitemOpen
  \bibfield  {author} {\bibinfo {author} {\bibfnamefont {J.~M.}\ \bibnamefont
  {Fink}}, \bibinfo {author} {\bibfnamefont {M.}~\bibnamefont {G\"oppl}},
  \bibinfo {author} {\bibfnamefont {M.}~\bibnamefont {Baur}}, \bibinfo {author}
  {\bibfnamefont {R.}~\bibnamefont {Bianchetti}}, \bibinfo {author}
  {\bibfnamefont {P.~J.}\ \bibnamefont {Leek}}, \bibinfo {author}
  {\bibfnamefont {A.}~\bibnamefont {Blais}}, \ and\ \bibinfo {author}
  {\bibfnamefont {A.}~\bibnamefont {Wallraff}},\ }\href@noop {} {\bibfield
  {journal} {\bibinfo  {journal} {Nature}\ }\textbf {\bibinfo {volume} {454}},\
  \bibinfo {pages} {315} (\bibinfo {year} {2008})}\BibitemShut {NoStop}%
\bibitem [{\citenamefont {Fink}\ \emph {et~al.}(2009)\citenamefont {Fink},
  \citenamefont {Bianchetti}, \citenamefont {Baur}, \citenamefont {G\"oppl},
  \citenamefont {Steffen}, \citenamefont {Filipp}, \citenamefont {Leek},
  \citenamefont {Blais},\ and\ \citenamefont {Wallraff}}]{Fink:2009}%
  \BibitemOpen
  \bibfield  {author} {\bibinfo {author} {\bibfnamefont {J.~M.}\ \bibnamefont
  {Fink}}, \bibinfo {author} {\bibfnamefont {R.}~\bibnamefont {Bianchetti}},
  \bibinfo {author} {\bibfnamefont {M.}~\bibnamefont {Baur}}, \bibinfo {author}
  {\bibfnamefont {M.}~\bibnamefont {G\"oppl}}, \bibinfo {author} {\bibfnamefont
  {L.}~\bibnamefont {Steffen}}, \bibinfo {author} {\bibfnamefont
  {S.}~\bibnamefont {Filipp}}, \bibinfo {author} {\bibfnamefont {P.~J.}\
  \bibnamefont {Leek}}, \bibinfo {author} {\bibfnamefont {A.}~\bibnamefont
  {Blais}}, \ and\ \bibinfo {author} {\bibfnamefont {A.}~\bibnamefont
  {Wallraff}},\ }\href {\doibase 10.1103/PhysRevLett.103.083601} {\bibfield
  {journal} {\bibinfo  {journal} {Phys. Rev. Lett.}\ }\textbf {\bibinfo
  {volume} {103}},\ \bibinfo {pages} {083601} (\bibinfo {year}
  {2009})}\BibitemShut {NoStop}%
\bibitem [{\citenamefont {Fragner}\ \emph {et~al.}(2008)\citenamefont
  {Fragner}, \citenamefont {G\"oppl}, \citenamefont {Fink}, \citenamefont
  {Baur}, \citenamefont {Bianchetti}, \citenamefont {Leek}, \citenamefont
  {Blais},\ and\ \citenamefont {Wallraff}}]{Fragner:2008}%
  \BibitemOpen
  \bibfield  {author} {\bibinfo {author} {\bibfnamefont {A.}~\bibnamefont
  {Fragner}}, \bibinfo {author} {\bibfnamefont {M.}~\bibnamefont {G\"oppl}},
  \bibinfo {author} {\bibfnamefont {J.~M.}\ \bibnamefont {Fink}}, \bibinfo
  {author} {\bibfnamefont {M.}~\bibnamefont {Baur}}, \bibinfo {author}
  {\bibfnamefont {R.}~\bibnamefont {Bianchetti}}, \bibinfo {author}
  {\bibfnamefont {P.~J.}\ \bibnamefont {Leek}}, \bibinfo {author}
  {\bibfnamefont {A.}~\bibnamefont {Blais}}, \ and\ \bibinfo {author}
  {\bibfnamefont {A.}~\bibnamefont {Wallraff}},\ }\href {\doibase
  10.1126/science.1164482} {\bibfield  {journal} {\bibinfo  {journal}
  {Science}\ }\textbf {\bibinfo {volume} {322}},\ \bibinfo {pages} {1357}
  (\bibinfo {year} {2008})}\BibitemShut {NoStop}%
\bibitem [{\citenamefont {Wilson}\ \emph {et~al.}(2011)\citenamefont {Wilson},
  \citenamefont {Johansson}, \citenamefont {Pourkabirian}, \citenamefont
  {Simoen}, \citenamefont {Johansson}, \citenamefont {Duty}, \citenamefont
  {Nori},\ and\ \citenamefont {Delsing}}]{wilson:2011}%
  \BibitemOpen
  \bibfield  {author} {\bibinfo {author} {\bibfnamefont {C.~M.}\ \bibnamefont
  {Wilson}}, \bibinfo {author} {\bibfnamefont {G.}~\bibnamefont {Johansson}},
  \bibinfo {author} {\bibfnamefont {A.}~\bibnamefont {Pourkabirian}}, \bibinfo
  {author} {\bibfnamefont {M.}~\bibnamefont {Simoen}}, \bibinfo {author}
  {\bibfnamefont {J.~R.}\ \bibnamefont {Johansson}}, \bibinfo {author}
  {\bibfnamefont {T.}~\bibnamefont {Duty}}, \bibinfo {author} {\bibfnamefont
  {F.}~\bibnamefont {Nori}}, \ and\ \bibinfo {author} {\bibfnamefont
  {P.}~\bibnamefont {Delsing}},\ }\href@noop {} {\bibfield  {journal} {\bibinfo
   {journal} {Nature}\ }\textbf {\bibinfo {volume} {479}},\ \bibinfo {pages}
  {376} (\bibinfo {year} {2011})}\BibitemShut {NoStop}%
\bibitem [{\citenamefont {L\"ahteenm\"aki}\ \emph {et~al.}(2013)\citenamefont
  {L\"ahteenm\"aki}, \citenamefont {Paraoanu}, \citenamefont {Hassel},\ and\
  \citenamefont {Hakonen}}]{hakonen:2012}%
  \BibitemOpen
  \bibfield  {author} {\bibinfo {author} {\bibfnamefont {P.}~\bibnamefont
  {L\"ahteenm\"aki}}, \bibinfo {author} {\bibfnamefont {G.~S.}\ \bibnamefont
  {Paraoanu}}, \bibinfo {author} {\bibfnamefont {J.}~\bibnamefont {Hassel}}, \
  and\ \bibinfo {author} {\bibfnamefont {P.~J.}\ \bibnamefont {Hakonen}},\
  }\href@noop {} {\bibfield  {journal} {\bibinfo  {journal} {PNAS}\ }\textbf
  {\bibinfo {volume} {110}},\ \bibinfo {pages} {4234} (\bibinfo {year}
  {2013})}\BibitemShut {NoStop}%
\bibitem [{\citenamefont {Myatt}\ \emph {et~al.}(2000)\citenamefont {Myatt},
  \citenamefont {King}, \citenamefont {Turchette}, \citenamefont {Sackett},
  \citenamefont {Kielpinski}, \citenamefont {Itano}, \citenamefont {Monroe},\
  and\ \citenamefont {Wineland}}]{Myatt:2000}%
  \BibitemOpen
  \bibfield  {author} {\bibinfo {author} {\bibfnamefont {C.~J.}\ \bibnamefont
  {Myatt}}, \bibinfo {author} {\bibfnamefont {B.~E.}\ \bibnamefont {King}},
  \bibinfo {author} {\bibfnamefont {Q.~A.}\ \bibnamefont {Turchette}}, \bibinfo
  {author} {\bibfnamefont {C.~A.}\ \bibnamefont {Sackett}}, \bibinfo {author}
  {\bibfnamefont {D.}~\bibnamefont {Kielpinski}}, \bibinfo {author}
  {\bibfnamefont {W.~M.}\ \bibnamefont {Itano}}, \bibinfo {author}
  {\bibfnamefont {C.}~\bibnamefont {Monroe}}, \ and\ \bibinfo {author}
  {\bibfnamefont {D.~J.}\ \bibnamefont {Wineland}},\ }\href@noop {} {\bibfield
  {journal} {\bibinfo  {journal} {Nature}\ }\textbf {\bibinfo {volume} {403}},\
  \bibinfo {pages} {269} (\bibinfo {year} {2000})}\BibitemShut {NoStop}%
\bibitem [{\citenamefont {Kielpinski}\ \emph {et~al.}(2001)\citenamefont
  {Kielpinski}, \citenamefont {Meyer}, \citenamefont {Rowe}, \citenamefont
  {Sackett}, \citenamefont {Itano}, \citenamefont {Monroe},\ and\ \citenamefont
  {Wineland}}]{Kielpinski:2001}%
  \BibitemOpen
  \bibfield  {author} {\bibinfo {author} {\bibfnamefont {D.}~\bibnamefont
  {Kielpinski}}, \bibinfo {author} {\bibfnamefont {V.}~\bibnamefont {Meyer}},
  \bibinfo {author} {\bibfnamefont {M.~A.}\ \bibnamefont {Rowe}}, \bibinfo
  {author} {\bibfnamefont {C.~A.}\ \bibnamefont {Sackett}}, \bibinfo {author}
  {\bibfnamefont {W.~M.}\ \bibnamefont {Itano}}, \bibinfo {author}
  {\bibfnamefont {C.}~\bibnamefont {Monroe}}, \ and\ \bibinfo {author}
  {\bibfnamefont {D.~J.}\ \bibnamefont {Wineland}},\ }\href@noop {} {\bibfield
  {journal} {\bibinfo  {journal} {Science}\ }\textbf {\bibinfo {volume}
  {291}},\ \bibinfo {pages} {1013} (\bibinfo {year} {2001})}\BibitemShut
  {NoStop}%
\bibitem [{\citenamefont {Gramich}\ \emph {et~al.}(2011)\citenamefont
  {Gramich}, \citenamefont {Solinas}, \citenamefont {M\"ott\"onen},
  \citenamefont {Pekola},\ and\ \citenamefont {Ankerhold}}]{Gramich:2011}%
  \BibitemOpen
  \bibfield  {author} {\bibinfo {author} {\bibfnamefont {V.}~\bibnamefont
  {Gramich}}, \bibinfo {author} {\bibfnamefont {P.}~\bibnamefont {Solinas}},
  \bibinfo {author} {\bibfnamefont {M.}~\bibnamefont {M\"ott\"onen}}, \bibinfo
  {author} {\bibfnamefont {J.~P.}\ \bibnamefont {Pekola}}, \ and\ \bibinfo
  {author} {\bibfnamefont {J.}~\bibnamefont {Ankerhold}},\ }\href {\doibase
  10.1103/PhysRevA.84.052103} {\bibfield  {journal} {\bibinfo  {journal} {Phys.
  Rev. A}\ }\textbf {\bibinfo {volume} {84}},\ \bibinfo {pages} {052103}
  (\bibinfo {year} {2011})}\BibitemShut {NoStop}%
\bibitem [{\citenamefont {Solinas}\ \emph {et~al.}(2012)\citenamefont
  {Solinas}, \citenamefont {M\"ott\"onen}, \citenamefont {Salmilehto},\ and\
  \citenamefont {Pekola}}]{Solinas:2012}%
  \BibitemOpen
  \bibfield  {author} {\bibinfo {author} {\bibfnamefont {P.}~\bibnamefont
  {Solinas}}, \bibinfo {author} {\bibfnamefont {M.}~\bibnamefont
  {M\"ott\"onen}}, \bibinfo {author} {\bibfnamefont {J.}~\bibnamefont
  {Salmilehto}}, \ and\ \bibinfo {author} {\bibfnamefont {J.~P.}\ \bibnamefont
  {Pekola}},\ }\href {\doibase 10.1103/PhysRevB.85.024527} {\bibfield
  {journal} {\bibinfo  {journal} {Phys. Rev. B}\ }\textbf {\bibinfo {volume}
  {85}},\ \bibinfo {pages} {024527} (\bibinfo {year} {2012})}\BibitemShut
  {NoStop}%
\bibitem [{\citenamefont {Murch}\ \emph {et~al.}(2012)\citenamefont {Murch},
  \citenamefont {Vool}, \citenamefont {Zhou}, \citenamefont {Weber},
  \citenamefont {Girvin},\ and\ \citenamefont {Siddiqi}}]{Murch:2012}%
  \BibitemOpen
  \bibfield  {author} {\bibinfo {author} {\bibfnamefont {K.~W.}\ \bibnamefont
  {Murch}}, \bibinfo {author} {\bibfnamefont {U.}~\bibnamefont {Vool}},
  \bibinfo {author} {\bibfnamefont {D.}~\bibnamefont {Zhou}}, \bibinfo {author}
  {\bibfnamefont {S.~J.}\ \bibnamefont {Weber}}, \bibinfo {author}
  {\bibfnamefont {S.~M.}\ \bibnamefont {Girvin}}, \ and\ \bibinfo {author}
  {\bibfnamefont {I.}~\bibnamefont {Siddiqi}},\ }\href@noop {} {\bibfield
  {journal} {\bibinfo  {journal} {Phys. Rev. Lett.}\ }\textbf {\bibinfo
  {volume} {109}},\ \bibinfo {pages} {183602} (\bibinfo {year}
  {2012})}\BibitemShut {NoStop}%
\bibitem [{\citenamefont {Oliver}\ \emph {et~al.}(2005)\citenamefont {Oliver},
  \citenamefont {Yu}, \citenamefont {Lee}, \citenamefont {Berggren},
  \citenamefont {Levitov},\ and\ \citenamefont {Orlando}}]{Oliver:2005}%
  \BibitemOpen
  \bibfield  {author} {\bibinfo {author} {\bibfnamefont {W.~D.}\ \bibnamefont
  {Oliver}}, \bibinfo {author} {\bibfnamefont {Y.}~\bibnamefont {Yu}}, \bibinfo
  {author} {\bibfnamefont {J.~C.}\ \bibnamefont {Lee}}, \bibinfo {author}
  {\bibfnamefont {K.~K.}\ \bibnamefont {Berggren}}, \bibinfo {author}
  {\bibfnamefont {L.~S.}\ \bibnamefont {Levitov}}, \ and\ \bibinfo {author}
  {\bibfnamefont {T.~P.}\ \bibnamefont {Orlando}},\ }\href {\doibase
  10.1126/science.1119678} {\bibfield  {journal} {\bibinfo  {journal}
  {Science}\ }\textbf {\bibinfo {volume} {310}},\ \bibinfo {pages} {1653}
  (\bibinfo {year} {2005})}\BibitemShut {NoStop}%
\bibitem [{\citenamefont {Wilson}\ \emph {et~al.}(2007)\citenamefont {Wilson},
  \citenamefont {Duty}, \citenamefont {Persson}, \citenamefont {Sandberg},
  \citenamefont {Johansson},\ and\ \citenamefont {Delsing}}]{Wilson:2007}%
  \BibitemOpen
  \bibfield  {author} {\bibinfo {author} {\bibfnamefont {C.~M.}\ \bibnamefont
  {Wilson}}, \bibinfo {author} {\bibfnamefont {T.}~\bibnamefont {Duty}},
  \bibinfo {author} {\bibfnamefont {F.}~\bibnamefont {Persson}}, \bibinfo
  {author} {\bibfnamefont {M.}~\bibnamefont {Sandberg}}, \bibinfo {author}
  {\bibfnamefont {G.}~\bibnamefont {Johansson}}, \ and\ \bibinfo {author}
  {\bibfnamefont {P.}~\bibnamefont {Delsing}},\ }\href {\doibase
  10.1103/PhysRevLett.98.257003} {\bibfield  {journal} {\bibinfo  {journal}
  {Phys. Rev. Lett.}\ }\textbf {\bibinfo {volume} {98}},\ \bibinfo {pages}
  {257003} (\bibinfo {year} {2007})}\BibitemShut {NoStop}%
\bibitem [{\citenamefont {Izmalkov}\ \emph {et~al.}(2008)\citenamefont
  {Izmalkov}, \citenamefont {van~der Ploeg}, \citenamefont {Shevchenko},
  \citenamefont {Grajcar}, \citenamefont {Il'ichev}, \citenamefont {H\"ubner},
  \citenamefont {Omelyanchouk},\ and\ \citenamefont {Meyer}}]{izmalkov:2008}%
  \BibitemOpen
  \bibfield  {author} {\bibinfo {author} {\bibfnamefont {A.}~\bibnamefont
  {Izmalkov}}, \bibinfo {author} {\bibfnamefont {S.~H.~W.}\ \bibnamefont
  {van~der Ploeg}}, \bibinfo {author} {\bibfnamefont {S.~N.}\ \bibnamefont
  {Shevchenko}}, \bibinfo {author} {\bibfnamefont {M.}~\bibnamefont {Grajcar}},
  \bibinfo {author} {\bibfnamefont {E.}~\bibnamefont {Il'ichev}}, \bibinfo
  {author} {\bibfnamefont {U.}~\bibnamefont {H\"ubner}}, \bibinfo {author}
  {\bibfnamefont {A.~N.}\ \bibnamefont {Omelyanchouk}}, \ and\ \bibinfo
  {author} {\bibfnamefont {H.-G.}\ \bibnamefont {Meyer}},\ }\href {\doibase
  10.1103/PhysRevLett.101.017003} {\bibfield  {journal} {\bibinfo  {journal}
  {Phys. Rev. Lett.}\ }\textbf {\bibinfo {volume} {101}},\ \bibinfo {pages}
  {017003} (\bibinfo {year} {2008})}\BibitemShut {NoStop}%
\bibitem [{\citenamefont {Gasparinetti}\ \emph {et~al.}(2013)\citenamefont
  {Gasparinetti}, \citenamefont {Solinas}, \citenamefont {Pugnetti},
  \citenamefont {Fazio},\ and\ \citenamefont {Pekola}}]{Gasparinetti:2013}%
  \BibitemOpen
  \bibfield  {author} {\bibinfo {author} {\bibfnamefont {S.}~\bibnamefont
  {Gasparinetti}}, \bibinfo {author} {\bibfnamefont {P.}~\bibnamefont
  {Solinas}}, \bibinfo {author} {\bibfnamefont {S.}~\bibnamefont {Pugnetti}},
  \bibinfo {author} {\bibfnamefont {R.}~\bibnamefont {Fazio}}, \ and\ \bibinfo
  {author} {\bibfnamefont {J.~P.}\ \bibnamefont {Pekola}},\ }\href {\doibase
  10.1103/PhysRevLett.110.150403} {\bibfield  {journal} {\bibinfo  {journal}
  {Phys. Rev. Lett.}\ }\textbf {\bibinfo {volume} {110}},\ \bibinfo {pages}
  {150403} (\bibinfo {year} {2013})}\BibitemShut {NoStop}%
\bibitem [{\citenamefont {Stace}\ \emph {et~al.}(2005)\citenamefont {Stace},
  \citenamefont {Doherty},\ and\ \citenamefont {Barrett}}]{stace:2005}%
  \BibitemOpen
  \bibfield  {author} {\bibinfo {author} {\bibfnamefont {T.~M.}\ \bibnamefont
  {Stace}}, \bibinfo {author} {\bibfnamefont {A.~C.}\ \bibnamefont {Doherty}},
  \ and\ \bibinfo {author} {\bibfnamefont {S.~D.}\ \bibnamefont {Barrett}},\
  }\href {\doibase 10.1103/PhysRevLett.95.106801} {\bibfield  {journal}
  {\bibinfo  {journal} {Phys. Rev. Lett.}\ }\textbf {\bibinfo {volume} {95}},\
  \bibinfo {pages} {106801} (\bibinfo {year} {2005})}\BibitemShut {NoStop}%
\bibitem [{\citenamefont {Stace}\ \emph {et~al.}(2013)\citenamefont {Stace},
  \citenamefont {Doherty},\ and\ \citenamefont {Reilly}}]{stace:2013}%
  \BibitemOpen
  \bibfield  {author} {\bibinfo {author} {\bibfnamefont {T.~M.}\ \bibnamefont
  {Stace}}, \bibinfo {author} {\bibfnamefont {A.~C.}\ \bibnamefont {Doherty}},
  \ and\ \bibinfo {author} {\bibfnamefont {D.~J.}\ \bibnamefont {Reilly}},\
  }\href {\doibase 10.1103/PhysRevLett.111.180602} {\bibfield  {journal}
  {\bibinfo  {journal} {Phys. Rev. Lett.}\ }\textbf {\bibinfo {volume} {111}},\
  \bibinfo {pages} {180602} (\bibinfo {year} {2013})}\BibitemShut {NoStop}%
\bibitem [{\citenamefont {Jentschura}\ \emph {et~al.}(2003)\citenamefont
  {Jentschura}, \citenamefont {Evers}, \citenamefont {Haas},\ and\
  \citenamefont {Keitel}}]{jentschura:2003}%
  \BibitemOpen
  \bibfield  {author} {\bibinfo {author} {\bibfnamefont {U.~D.}\ \bibnamefont
  {Jentschura}}, \bibinfo {author} {\bibfnamefont {J.}~\bibnamefont {Evers}},
  \bibinfo {author} {\bibfnamefont {M.}~\bibnamefont {Haas}}, \ and\ \bibinfo
  {author} {\bibfnamefont {C.~H.}\ \bibnamefont {Keitel}},\ }\href@noop {}
  {\bibfield  {journal} {\bibinfo  {journal} {Phys. Rev. Lett.}\ }\textbf
  {\bibinfo {volume} {91}},\ \bibinfo {pages} {253601} (\bibinfo {year}
  {2003})}\BibitemShut {NoStop}%
\bibitem [{\citenamefont {Ramsay}\ \emph {et~al.}(2010)\citenamefont {Ramsay},
  \citenamefont {Godden}, \citenamefont {Boyle}, \citenamefont {Gauger},
  \citenamefont {Nazir}, \citenamefont {Lovett}, \citenamefont {Fox},\ and\
  \citenamefont {Skolnick}}]{Ramsay:2010}%
  \BibitemOpen
  \bibfield  {author} {\bibinfo {author} {\bibfnamefont {A.~J.}\ \bibnamefont
  {Ramsay}}, \bibinfo {author} {\bibfnamefont {T.~M.}\ \bibnamefont {Godden}},
  \bibinfo {author} {\bibfnamefont {S.~J.}\ \bibnamefont {Boyle}}, \bibinfo
  {author} {\bibfnamefont {E.~M.}\ \bibnamefont {Gauger}}, \bibinfo {author}
  {\bibfnamefont {A.}~\bibnamefont {Nazir}}, \bibinfo {author} {\bibfnamefont
  {B.~W.}\ \bibnamefont {Lovett}}, \bibinfo {author} {\bibfnamefont {A.~M.}\
  \bibnamefont {Fox}}, \ and\ \bibinfo {author} {\bibfnamefont {M.~S.}\
  \bibnamefont {Skolnick}},\ }\href {\doibase 10.1103/PhysRevLett.105.177402}
  {\bibfield  {journal} {\bibinfo  {journal} {Phys. Rev. Lett.}\ }\textbf
  {\bibinfo {volume} {105}},\ \bibinfo {pages} {177402} (\bibinfo {year}
  {2010})}\BibitemShut {NoStop}%
\bibitem [{\citenamefont {Colless}\ \emph {et~al.}(2013)\citenamefont
  {Colless}, \citenamefont {Croot}, \citenamefont {Stace}, \citenamefont
  {Doherty}, \citenamefont {Barrett}, \citenamefont {Lu}, \citenamefont
  {Gossard},\ and\ \citenamefont {Reilly}}]{colless:2013}%
  \BibitemOpen
  \bibfield  {author} {\bibinfo {author} {\bibfnamefont {J.~I.}\ \bibnamefont
  {Colless}}, \bibinfo {author} {\bibfnamefont {X.~G.}\ \bibnamefont {Croot}},
  \bibinfo {author} {\bibfnamefont {T.~M.}\ \bibnamefont {Stace}}, \bibinfo
  {author} {\bibfnamefont {A.~C.}\ \bibnamefont {Doherty}}, \bibinfo {author}
  {\bibfnamefont {S.~D.}\ \bibnamefont {Barrett}}, \bibinfo {author}
  {\bibfnamefont {H.}~\bibnamefont {Lu}}, \bibinfo {author} {\bibfnamefont
  {A.~C.}\ \bibnamefont {Gossard}}, \ and\ \bibinfo {author} {\bibfnamefont
  {D.~J.}\ \bibnamefont {Reilly}},\ }\href@noop {} {\bibfield  {journal}
  {\bibinfo  {journal} {arXiv:1305.5982v1 [cond-mat.mes-hall]}\ } (\bibinfo
  {year} {2013})}\BibitemShut {NoStop}%
\bibitem [{\citenamefont {Niskanen}\ \emph {et~al.}(2003)\citenamefont
  {Niskanen}, \citenamefont {Pekola},\ and\ \citenamefont
  {Sepp\"a}}]{niskanen:2003}%
  \BibitemOpen
  \bibfield  {author} {\bibinfo {author} {\bibfnamefont {A.~O.}\ \bibnamefont
  {Niskanen}}, \bibinfo {author} {\bibfnamefont {J.~P.}\ \bibnamefont
  {Pekola}}, \ and\ \bibinfo {author} {\bibfnamefont {H.}~\bibnamefont
  {Sepp\"a}},\ }\href {\doibase 10.1103/PhysRevLett.91.177003} {\bibfield
  {journal} {\bibinfo  {journal} {Phys. Rev. Lett.}\ }\textbf {\bibinfo
  {volume} {91}},\ \bibinfo {pages} {177003} (\bibinfo {year}
  {2003})}\BibitemShut {NoStop}%
\bibitem [{\citenamefont {M\"ott\"onen}\ \emph {et~al.}(2008)\citenamefont
  {M\"ott\"onen}, \citenamefont {Vartiainen},\ and\ \citenamefont
  {Pekola}}]{Mottonen:2008}%
  \BibitemOpen
  \bibfield  {author} {\bibinfo {author} {\bibfnamefont {M.}~\bibnamefont
  {M\"ott\"onen}}, \bibinfo {author} {\bibfnamefont {J.~J.}\ \bibnamefont
  {Vartiainen}}, \ and\ \bibinfo {author} {\bibfnamefont {J.~P.}\ \bibnamefont
  {Pekola}},\ }\href {\doibase 10.1103/PhysRevLett.100.177201} {\bibfield
  {journal} {\bibinfo  {journal} {Phys. Rev. Lett.}\ }\textbf {\bibinfo
  {volume} {100}},\ \bibinfo {pages} {177201} (\bibinfo {year}
  {2008})}\BibitemShut {NoStop}%
\bibitem [{\citenamefont {Gasparinetti}\ \emph {et~al.}(2011)\citenamefont
  {Gasparinetti}, \citenamefont {Solinas},\ and\ \citenamefont
  {Pekola}}]{Gasparinetti:2011}%
  \BibitemOpen
  \bibfield  {author} {\bibinfo {author} {\bibfnamefont {S.}~\bibnamefont
  {Gasparinetti}}, \bibinfo {author} {\bibfnamefont {P.}~\bibnamefont
  {Solinas}}, \ and\ \bibinfo {author} {\bibfnamefont {J.~P.}\ \bibnamefont
  {Pekola}},\ }\href {\doibase 10.1103/PhysRevLett.107.207002} {\bibfield
  {journal} {\bibinfo  {journal} {Phys. Rev. Lett.}\ }\textbf {\bibinfo
  {volume} {107}},\ \bibinfo {pages} {207002} (\bibinfo {year}
  {2011})}\BibitemShut {NoStop}%
\bibitem [{\citenamefont {Gasparinetti}\ \emph {et~al.}(2012)\citenamefont
  {Gasparinetti}, \citenamefont {Solinas}, \citenamefont {Yoon},\ and\
  \citenamefont {Pekola}}]{Gasparinetti:2012}%
  \BibitemOpen
  \bibfield  {author} {\bibinfo {author} {\bibfnamefont {S.}~\bibnamefont
  {Gasparinetti}}, \bibinfo {author} {\bibfnamefont {P.}~\bibnamefont
  {Solinas}}, \bibinfo {author} {\bibfnamefont {Y.}~\bibnamefont {Yoon}}, \
  and\ \bibinfo {author} {\bibfnamefont {J.~P.}\ \bibnamefont {Pekola}},\
  }\href {\doibase 10.1103/PhysRevB.86.060502} {\bibfield  {journal} {\bibinfo
  {journal} {Phys. Rev. B}\ }\textbf {\bibinfo {volume} {86}},\ \bibinfo
  {pages} {060502(R)} (\bibinfo {year} {2012})}\BibitemShut {NoStop}%
\bibitem [{\citenamefont {Grifoni}\ and\ \citenamefont
  {H\"anggi}(1998)}]{Grifoni:1998}%
  \BibitemOpen
  \bibfield  {author} {\bibinfo {author} {\bibfnamefont {M.}~\bibnamefont
  {Grifoni}}\ and\ \bibinfo {author} {\bibfnamefont {P.}~\bibnamefont
  {H\"anggi}},\ }\href {\doibase
  http://dx.doi.org/10.1016/S0370-1573(98)00022-2} {\bibfield  {journal}
  {\bibinfo  {journal} {Physics Reports}\ }\textbf {\bibinfo {volume} {304}},\
  \bibinfo {pages} {229 } (\bibinfo {year} {1998})}\BibitemShut {NoStop}%
\bibitem [{epa()}]{epaps}%
  \BibitemOpen
  \href@noop {} {}\bibinfo {note} {See Supplemental at XXX for details on the
  LS and the Floquet master equation}\BibitemShut {NoStop}%
\bibitem [{\citenamefont {Russomanno}\ \emph {et~al.}(2011)\citenamefont
  {Russomanno}, \citenamefont {Pugnetti}, \citenamefont {Brosco},\ and\
  \citenamefont {Fazio}}]{Russomanno:2011}%
  \BibitemOpen
  \bibfield  {author} {\bibinfo {author} {\bibfnamefont {A.}~\bibnamefont
  {Russomanno}}, \bibinfo {author} {\bibfnamefont {S.}~\bibnamefont
  {Pugnetti}}, \bibinfo {author} {\bibfnamefont {V.}~\bibnamefont {Brosco}}, \
  and\ \bibinfo {author} {\bibfnamefont {R.}~\bibnamefont {Fazio}},\ }\href
  {\doibase 10.1103/PhysRevB.83.214508} {\bibfield  {journal} {\bibinfo
  {journal} {Phys. Rev. B}\ }\textbf {\bibinfo {volume} {83}},\ \bibinfo
  {pages} {214508} (\bibinfo {year} {2011})}\BibitemShut {NoStop}%
\bibitem [{\citenamefont {Scully}\ and\ \citenamefont
  {Zubairy}(1997)}]{scully:1997}%
  \BibitemOpen
  \bibfield  {author} {\bibinfo {author} {\bibfnamefont {M.~O.}\ \bibnamefont
  {Scully}}\ and\ \bibinfo {author} {\bibfnamefont {M.~S.}\ \bibnamefont
  {Zubairy}},\ }\href@noop {} {\emph {\bibinfo {title} {{Quantum Optics}}}}\
  (\bibinfo  {publisher} {Cambridge University Press},\ \bibinfo {year}
  {1997})\BibitemShut {NoStop}%
\bibitem [{\citenamefont {Wilson}\ \emph {et~al.}(2010)\citenamefont {Wilson},
  \citenamefont {Johansson}, \citenamefont {Duty}, \citenamefont {Persson},
  \citenamefont {Sandberg},\ and\ \citenamefont {Delsing}}]{wilson:2010}%
  \BibitemOpen
  \bibfield  {author} {\bibinfo {author} {\bibfnamefont {C.~M.}\ \bibnamefont
  {Wilson}}, \bibinfo {author} {\bibfnamefont {G.}~\bibnamefont {Johansson}},
  \bibinfo {author} {\bibfnamefont {T.}~\bibnamefont {Duty}}, \bibinfo {author}
  {\bibfnamefont {F.}~\bibnamefont {Persson}}, \bibinfo {author} {\bibfnamefont
  {M.}~\bibnamefont {Sandberg}}, \ and\ \bibinfo {author} {\bibfnamefont
  {P.}~\bibnamefont {Delsing}},\ }\href {\doibase 10.1103/PhysRevB.81.024520}
  {\bibfield  {journal} {\bibinfo  {journal} {Phys. Rev. B}\ }\textbf {\bibinfo
  {volume} {81}},\ \bibinfo {pages} {024520} (\bibinfo {year}
  {2010})}\BibitemShut {NoStop}%
\bibitem [{\citenamefont {Silveri}\ \emph {et~al.}(2013)\citenamefont
  {Silveri}, \citenamefont {Tuorila}, \citenamefont {Kemppainen},\ and\
  \citenamefont {Thuneberg}}]{Silveri:2013}%
  \BibitemOpen
  \bibfield  {author} {\bibinfo {author} {\bibfnamefont {M.}~\bibnamefont
  {Silveri}}, \bibinfo {author} {\bibfnamefont {J.}~\bibnamefont {Tuorila}},
  \bibinfo {author} {\bibfnamefont {M.}~\bibnamefont {Kemppainen}}, \ and\
  \bibinfo {author} {\bibfnamefont {E.}~\bibnamefont {Thuneberg}},\ }\href
  {\doibase 10.1103/PhysRevB.87.134505} {\bibfield  {journal} {\bibinfo
  {journal} {Phys. Rev. B}\ }\textbf {\bibinfo {volume} {87}},\ \bibinfo
  {pages} {134505} (\bibinfo {year} {2013})}\BibitemShut {NoStop}%
\bibitem [{\citenamefont {Niskanen}\ \emph {et~al.}(2005)\citenamefont
  {Niskanen}, \citenamefont {Kivioja}, \citenamefont {Sepp\"a},\ and\
  \citenamefont {Pekola}}]{Niskanen:2005}%
  \BibitemOpen
  \bibfield  {author} {\bibinfo {author} {\bibfnamefont {A.~O.}\ \bibnamefont
  {Niskanen}}, \bibinfo {author} {\bibfnamefont {J.~M.}\ \bibnamefont
  {Kivioja}}, \bibinfo {author} {\bibfnamefont {H.}~\bibnamefont {Sepp\"a}}, \
  and\ \bibinfo {author} {\bibfnamefont {J.~P.}\ \bibnamefont {Pekola}},\
  }\href {\doibase 10.1103/PhysRevB.71.012513} {\bibfield  {journal} {\bibinfo
  {journal} {Phys. Rev. B}\ }\textbf {\bibinfo {volume} {71}},\ \bibinfo
  {pages} {012513} (\bibinfo {year} {2005})}\BibitemShut {NoStop}%
\bibitem [{\citenamefont {Solinas}\ \emph {et~al.}(2010)\citenamefont
  {Solinas}, \citenamefont {M\"ott\"onen}, \citenamefont {Salmilehto},\ and\
  \citenamefont {Pekola}}]{Solinas:2010}%
  \BibitemOpen
  \bibfield  {author} {\bibinfo {author} {\bibfnamefont {P.}~\bibnamefont
  {Solinas}}, \bibinfo {author} {\bibfnamefont {M.}~\bibnamefont
  {M\"ott\"onen}}, \bibinfo {author} {\bibfnamefont {J.}~\bibnamefont
  {Salmilehto}}, \ and\ \bibinfo {author} {\bibfnamefont {J.~P.}\ \bibnamefont
  {Pekola}},\ }\href {\doibase 10.1103/PhysRevB.82.134517} {\bibfield
  {journal} {\bibinfo  {journal} {Phys. Rev. B}\ }\textbf {\bibinfo {volume}
  {82}},\ \bibinfo {pages} {134517} (\bibinfo {year} {2010})}\BibitemShut
  {NoStop}%
\end{thebibliography}

\begin{thebibliography}{8}%
\makeatletter
\providecommand \@ifxundefined [1]{%
 \@ifx{#1\undefined}
}%
\providecommand \@ifnum [1]{%
 \ifnum #1\expandafter \@firstoftwo
 \else \expandafter \@secondoftwo
 \fi
}%
\providecommand \@ifx [1]{%
 \ifx #1\expandafter \@firstoftwo
 \else \expandafter \@secondoftwo
 \fi
}%
\providecommand \natexlab [1]{#1}%
\providecommand \enquote  [1]{``#1''}%
\providecommand \bibnamefont  [1]{#1}%
\providecommand \bibfnamefont [1]{#1}%
\providecommand \citenamefont [1]{#1}%
\providecommand \href@noop [0]{\@secondoftwo}%
\providecommand \href [0]{\begingroup \@sanitize@url \@href}%
\providecommand \@href[1]{\@@startlink{#1}\@@href}%
\providecommand \@@href[1]{\endgroup#1\@@endlink}%
\providecommand \@sanitize@url [0]{\catcode `\\12\catcode `\$12\catcode
  `\&12\catcode `\#12\catcode `\^12\catcode `\_12\catcode `\%12\relax}%
\providecommand \@@startlink[1]{}%
\providecommand \@@endlink[0]{}%
\providecommand \url  [0]{\begingroup\@sanitize@url \@url }%
\providecommand \@url [1]{\endgroup\@href {#1}{\urlprefix }}%
\providecommand \urlprefix  [0]{URL }%
\providecommand \Eprint [0]{\href }%
\providecommand \doibase [0]{http://dx.doi.org/}%
\providecommand \selectlanguage [0]{\@gobble}%
\providecommand \bibinfo  [0]{\@secondoftwo}%
\providecommand \bibfield  [0]{\@secondoftwo}%
\providecommand \translation [1]{[#1]}%
\providecommand \BibitemOpen [0]{}%
\providecommand \bibitemStop [0]{}%
\providecommand \bibitemNoStop [0]{.\EOS\space}%
\providecommand \EOS [0]{\spacefactor3000\relax}%
\providecommand \BibitemShut  [1]{\csname bibitem#1\endcsname}%
\let\auto@bib@innerbib\@empty
%</preamble>
\bibitem [{\citenamefont {Bl\"umel}\ \emph {et~al.}(1991)\citenamefont
  {Bl\"umel}, \citenamefont {Buchleitner}, \citenamefont {Graham},
  \citenamefont {Sirko}, \citenamefont {Smilansky},\ and\ \citenamefont
  {Walther}}]{blueml:1991}%
  \BibitemOpen
  \bibfield  {author} {\bibinfo {author} {\bibfnamefont {R.}~\bibnamefont
  {Bl\"umel}}, \bibinfo {author} {\bibfnamefont {A.}~\bibnamefont
  {Buchleitner}}, \bibinfo {author} {\bibfnamefont {R.}~\bibnamefont {Graham}},
  \bibinfo {author} {\bibfnamefont {L.}~\bibnamefont {Sirko}}, \bibinfo
  {author} {\bibfnamefont {U.}~\bibnamefont {Smilansky}}, \ and\ \bibinfo
  {author} {\bibfnamefont {H.}~\bibnamefont {Walther}},\ }\href {\doibase
  10.1103/PhysRevA.44.4521} {\bibfield  {journal} {\bibinfo  {journal} {Phys.
  Rev. A}\ }\textbf {\bibinfo {volume} {44}},\ \bibinfo {pages} {4521}
  (\bibinfo {year} {1991})}\BibitemShut {NoStop}%
\bibitem [{\citenamefont {Grifoni}\ and\ \citenamefont
  {H\"anggi}(1998)}]{Grifoni1:1998}%
  \BibitemOpen
  \bibfield  {author} {\bibinfo {author} {\bibfnamefont {M.}~\bibnamefont
  {Grifoni}}\ and\ \bibinfo {author} {\bibfnamefont {P.}~\bibnamefont
  {H\"anggi}},\ }\href {\doibase
  http://dx.doi.org/10.1016/S0370-1573(98)00022-2} {\bibfield  {journal}
  {\bibinfo  {journal} {Physics Reports}\ }\textbf {\bibinfo {volume} {304}},\
  \bibinfo {pages} {229 } (\bibinfo {year} {1998})}\BibitemShut {NoStop}%
\bibitem [{\citenamefont {Gasparinetti}\ \emph {et~al.}(2013)\citenamefont
  {Gasparinetti}, \citenamefont {Solinas}, \citenamefont {Pugnetti},
  \citenamefont {Fazio},\ and\ \citenamefont {Pekola}}]{Gasparinetti1:2013}%
  \BibitemOpen
  \bibfield  {author} {\bibinfo {author} {\bibfnamefont {S.}~\bibnamefont
  {Gasparinetti}}, \bibinfo {author} {\bibfnamefont {P.}~\bibnamefont
  {Solinas}}, \bibinfo {author} {\bibfnamefont {S.}~\bibnamefont {Pugnetti}},
  \bibinfo {author} {\bibfnamefont {R.}~\bibnamefont {Fazio}}, \ and\ \bibinfo
  {author} {\bibfnamefont {J.~P.}\ \bibnamefont {Pekola}},\ }\href {\doibase
  10.1103/PhysRevLett.110.150403} {\bibfield  {journal} {\bibinfo  {journal}
  {Phys. Rev. Lett.}\ }\textbf {\bibinfo {volume} {110}},\ \bibinfo {pages}
  {150403} (\bibinfo {year} {2013})}\BibitemShut {NoStop}%
\bibitem [{\citenamefont {Santana}\ \emph {et~al.}(2001)\citenamefont
  {Santana}, \citenamefont {Gomez~Llorente},\ and\ \citenamefont
  {Delgado}}]{santana:2001}%
  \BibitemOpen
  \bibfield  {author} {\bibinfo {author} {\bibfnamefont {A.}~\bibnamefont
  {Santana}}, \bibinfo {author} {\bibfnamefont {J.~M.}\ \bibnamefont
  {Gomez~Llorente}}, \ and\ \bibinfo {author} {\bibfnamefont {V.}~\bibnamefont
  {Delgado}},\ }\href@noop {} {\bibfield  {journal} {\bibinfo  {journal} {J.
  Phys. B: At. Mol. Opt. Phys.}\ }\textbf {\bibinfo {volume} {34}},\ \bibinfo
  {pages} {2371} (\bibinfo {year} {2001})}\BibitemShut {NoStop}%
\bibitem [{\citenamefont {Oliver}\ \emph {et~al.}(2005)\citenamefont {Oliver},
  \citenamefont {Yu}, \citenamefont {Lee}, \citenamefont {Berggren},
  \citenamefont {Levitov},\ and\ \citenamefont {Orlando}}]{Oliver1:2005}%
  \BibitemOpen
  \bibfield  {author} {\bibinfo {author} {\bibfnamefont {W.~D.}\ \bibnamefont
  {Oliver}}, \bibinfo {author} {\bibfnamefont {Y.}~\bibnamefont {Yu}}, \bibinfo
  {author} {\bibfnamefont {J.~C.}\ \bibnamefont {Lee}}, \bibinfo {author}
  {\bibfnamefont {K.~K.}\ \bibnamefont {Berggren}}, \bibinfo {author}
  {\bibfnamefont {L.~S.}\ \bibnamefont {Levitov}}, \ and\ \bibinfo {author}
  {\bibfnamefont {T.~P.}\ \bibnamefont {Orlando}},\ }\href {\doibase
  10.1126/science.1119678} {\bibfield  {journal} {\bibinfo  {journal}
  {Science}\ }\textbf {\bibinfo {volume} {310}},\ \bibinfo {pages} {1653}
  (\bibinfo {year} {2005})}\BibitemShut {NoStop}%
\bibitem [{\citenamefont {Wilson}\ \emph {et~al.}(2007)\citenamefont {Wilson},
  \citenamefont {Duty}, \citenamefont {Persson}, \citenamefont {Sandberg},
  \citenamefont {Johansson},\ and\ \citenamefont {Delsing}}]{Wilson1:2007}%
  \BibitemOpen
  \bibfield  {author} {\bibinfo {author} {\bibfnamefont {C.~M.}\ \bibnamefont
  {Wilson}}, \bibinfo {author} {\bibfnamefont {T.}~\bibnamefont {Duty}},
  \bibinfo {author} {\bibfnamefont {F.}~\bibnamefont {Persson}}, \bibinfo
  {author} {\bibfnamefont {M.}~\bibnamefont {Sandberg}}, \bibinfo {author}
  {\bibfnamefont {G.}~\bibnamefont {Johansson}}, \ and\ \bibinfo {author}
  {\bibfnamefont {P.}~\bibnamefont {Delsing}},\ }\href {\doibase
  10.1103/PhysRevLett.98.257003} {\bibfield  {journal} {\bibinfo  {journal}
  {Phys. Rev. Lett.}\ }\textbf {\bibinfo {volume} {98}},\ \bibinfo {pages}
  {257003} (\bibinfo {year} {2007})}\BibitemShut {NoStop}%
\bibitem [{\citenamefont {Wilson}\ \emph {et~al.}(2010)\citenamefont {Wilson},
  \citenamefont {Johansson}, \citenamefont {Duty}, \citenamefont {Persson},
  \citenamefont {Sandberg},\ and\ \citenamefont {Delsing}}]{Wilson1:2010}%
  \BibitemOpen
  \bibfield  {author} {\bibinfo {author} {\bibfnamefont {C.~M.}\ \bibnamefont
  {Wilson}}, \bibinfo {author} {\bibfnamefont {G.}~\bibnamefont {Johansson}},
  \bibinfo {author} {\bibfnamefont {T.}~\bibnamefont {Duty}}, \bibinfo {author}
  {\bibfnamefont {F.}~\bibnamefont {Persson}}, \bibinfo {author} {\bibfnamefont
  {M.}~\bibnamefont {Sandberg}}, \ and\ \bibinfo {author} {\bibfnamefont
  {P.}~\bibnamefont {Delsing}},\ }\href {\doibase 10.1103/PhysRevB.81.024520}
  {\bibfield  {journal} {\bibinfo  {journal} {Phys. Rev. B}\ }\textbf {\bibinfo
  {volume} {81}},\ \bibinfo {pages} {024520} (\bibinfo {year}
  {2010})}\BibitemShut {NoStop}%
\bibitem [{\citenamefont {Fragner}\ \emph {et~al.}(2008)\citenamefont
  {Fragner}, \citenamefont {G\"oppl}, \citenamefont {Fink}, \citenamefont
  {Baur}, \citenamefont {Bianchetti}, \citenamefont {Leek}, \citenamefont
  {Blais},\ and\ \citenamefont {Wallraff}}]{Fragner1:2008}%
  \BibitemOpen
  \bibfield  {author} {\bibinfo {author} {\bibfnamefont {A.}~\bibnamefont
  {Fragner}}, \bibinfo {author} {\bibfnamefont {M.}~\bibnamefont {G\"oppl}},
  \bibinfo {author} {\bibfnamefont {J.~M.}\ \bibnamefont {Fink}}, \bibinfo
  {author} {\bibfnamefont {M.}~\bibnamefont {Baur}}, \bibinfo {author}
  {\bibfnamefont {R.}~\bibnamefont {Bianchetti}}, \bibinfo {author}
  {\bibfnamefont {P.~J.}\ \bibnamefont {Leek}}, \bibinfo {author}
  {\bibfnamefont {A.}~\bibnamefont {Blais}}, \ and\ \bibinfo {author}
  {\bibfnamefont {A.}~\bibnamefont {Wallraff}},\ }\href {\doibase
  10.1126/science.1164482} {\bibfield  {journal} {\bibinfo  {journal}
  {Science}\ }\textbf {\bibinfo {volume} {322}},\ \bibinfo {pages} {1357}
  (\bibinfo {year} {2008})}\BibitemShut {NoStop}%
\end{thebibliography}


\begin{widetext}
\appendix
\newpage
\section{Supplemental Material}

\begin{small}
\noindent
In this supplemental material to our article `Lamb shift enhancement and detection in strongly driven superconducting circuits' we present further details about the derivation of the Lamb shift as well as the semiclassical Rabi model, its Lamb shift for a single mode reservoir, and its equivalence to experimental realizations in limiting cases.
\end{small} 

\section{Derivation of the Lamb shift}

To reveal the impact of the Lamb shift expression given in Eq.~(4) in the main text, we present here its full derivation, starting with the time-independent Floquet master equation (for further details follow the steps in, e.g., Refs.~\cite{blueml:1991, Grifoni1:1998, Gasparinetti1:2013}) written in the basis of Floquet modes $|\Phi_{\alpha}(t)\rangle = |\Phi_{\alpha}(t+\frac{2\pi}{\Omega})\rangle$ and in the Schr\"odinger picture
\begin{equation}
\dot{\rho}_{\alpha\beta}(t) = -i (\omega_{\alpha\beta}-\delta\omega_{\alpha\beta})\,\rho_{\alpha\beta}(t)+ \sum_{\gamma,\delta}\mathcal{R}_{\alpha\beta\gamma\delta}\,\rho_{\gamma\delta}(t)\,,
\label{eq:standardnotation}
\end{equation}
where
\begin{equation}
\mathcal{R}_{\alpha\beta\gamma\delta}=\Gamma^{+}_{\alpha\gamma\beta\delta} + \Gamma^{-}_{\alpha\gamma\beta\delta}-\delta_{\delta,\beta}\sum_{\mu}\Gamma^{+}_{\mu\gamma\mu\alpha}-\delta_{\gamma,\alpha}\sum_{\mu}\Gamma^{-}_{\mu\beta\mu\delta}
\label{eq:Redfieldtensor}
\end{equation}
represents the Redfield tensor and
\begin{equation}
\Gamma^{+}_{\alpha\beta\gamma\delta}=\frac{1}{\hbar}\sum_k \mathcal{S}(\Delta_{\alpha\beta,  k})X_{\alpha\beta,k}^{(r)}(X_{\gamma\delta,k}^{(r)})^{\ast}, \,\,\Gamma^{-}_{\alpha\beta\gamma\delta}=\frac{1}{\hbar}\sum_k \mathcal{S}(\Delta_{\gamma\delta,k})X_{\alpha\beta,k}^{(r)}(X_{\gamma\delta,k}^{(r)})^{\ast}
\label{eq:Gammarates}
\end{equation}
the effective relaxation and dephasing rates, respectively. A crucial point
in deriving the result \eqref{eq:standardnotation} is that we have performed the so-called partial secular approximation (PSA) ~\cite{Gasparinetti1:2013}. It consists in retaining all terms  which oscillate with $\epsilon_{\alpha}\neq \epsilon_{\beta}$ (in contrast to the usual rotating wave approximation, RWA) and neglecting only those with multiple integers $k_{\alpha},k_{\beta}$ of $\hbar\Omega$ where $k_{\alpha}\neq k_{\beta}$. The PSA imposes the constraint $\Omega\gg \eta\,|\omega_{\alpha\beta}|$ which is a much weaker condition than that for a full RWA which requires $|\omega_{\alpha\beta}|\gg \Gamma$ with a typical relaxation time scale $1/\Gamma$. For details see \cite{Grifoni1:1998, Gasparinetti1:2013}.

The Lamb shift contributions $\delta\omega_{\alpha\beta}$
arise from principal value terms via $\int_0^{\infty} dt\, {\rm e}^{i\omega t}=\pi\delta(\omega) + i\,\mathcal{P}(1/\omega)$ in the Floquet master equation in the basis of Floquet states $|\Psi_{\alpha}(t)\rangle$ following the procedure outlined in Refs.~\cite{Grifoni1:1998, Gasparinetti1:2013}. The matrix elements
\begin{equation}
 X_{\alpha\beta,k}^{(r)}= \frac{\Omega}{2\pi}\int_{0}^{2\pi/\Omega} dt\, {\rm e}^{-ik\Omega t}\langle \Phi_{\alpha}(t)|S(r)|\Phi_{\beta}(t)\rangle
  \end{equation}
  with the driving frequency $\Omega$ contain the system operator $S$ coupling to the reservoir and obey the symmetry relation $X_{\alpha\beta, k}^{(r)}=X^{\ast \;(r)}_{\beta\alpha,-k}$. The latter one helps to simplify the Lamb shift terms. Transition energies are given by
 \begin{equation}
 \hbar \Delta_{\alpha\beta,k} = \epsilon_{\alpha}-\epsilon_{\beta} + k\hbar\Omega
 \label{eq:transition}
 \end{equation}
with $\epsilon_{i}(t), i=\alpha, \beta$ playing the role of a dressed state energy.
The quantity $\omega_{\alpha\beta}=(\epsilon_{\alpha}-\epsilon_{\beta})/\hbar$ thereby expresses the Floquet quasienergy gap which is a non-dissipative contribution of the driven quantum system.
In the above formulas we have also used the abbreviation
\begin{eqnarray}
\mathcal{S}(\omega)=\theta(\omega)J(\omega)n_{\rm th}(\omega) + \theta(-\omega)J(-\omega)[n_{\rm th}(-\omega)+1]
\label{eq:Sterm}
\end{eqnarray}
containing the spectral bath density $J(\omega)$ in units of an energy. Here, $\theta(\omega)$ denotes the Heavyside function and $n_{\rm th}(\omega)$ is the usual Bose-Einstein distribution.

In the above derivation principal value corrections to the rates in Eq.~\eqref{eq:Gammarates} are neglected as they provide higher order corrections only. In the singular coupling regime to be of main interest here, the Lamb shift provides the leading order contribution. To evaluate it explicitly, one has to consider integrals of the form
\begin{equation}
G^{\pm}(\Delta_{\alpha\beta,k})=\mathcal{P}\int_0^{\infty} d\omega\, \frac{J(\omega)n_{\rm th}(\pm\omega)}{\omega - \Delta_{\alpha\beta,k}}\, .
\label{eq:PVintegral}
\end{equation}
This leads us to the following expression
\begin{eqnarray}
\delta\omega_{\alpha\beta}=\frac{1}{\pi\hbar}\sum_{\mu, k}\hspace{-0.3cm}&\bigl[|X_{\beta\mu,k}^{(r)}|^2\{G^+(\Delta_{\mu\beta,-k})+G^-(\Delta_{\beta\mu,k})\} \nonumber\\
&-|X_{\mu\alpha,k}^{(r)}|^2\{G^+(\Delta_{\mu\alpha,k})+G^-(\Delta_{\alpha\mu,-k})\}
\bigr]\,.
\label{eq:lscompact}
\end{eqnarray}
The above results apply to arbitrary spectral density of the bath $J(\omega)$. Now, for a TLS ($\mu=1,2$) at zero temperature (only zero-point fluctuations) $n_\beta(\omega>0)\to 0$, the $G^+$-contributions drop out. Applying the symmetry relations $X_{\alpha\beta, k}^{(r)}=X^{\ast \;(r)}_{\beta\alpha,-k}$ as well as $X_{11,k}^{(r)}=-X_{22,k}^{(r)}$ for a traceless noise operator (which is true for all combinations of Pauli matrices) and orthogonal Floquet modes at all times, only terms with different indices in the coupling matrix elements survive in (\ref{eq:lscompact}). This leads to Eq. (4) in the main text.

For an Ohmic-type distribution with exponential cut-off  $J(\omega)=\eta\hbar\omega \exp(-\omega/\omega_c)$ with a dimensionless coupling constant $\eta$ and a large cut-off frequency $\omega_c$, one obtains
\begin{equation}
G(\Delta_{\alpha\beta,k})\equiv G^-_{T=0}(\Delta_{\alpha\beta,k})=-\eta\hbar\omega_c+\eta\hbar\Delta_{\alpha\beta,k}\, {\rm e}^{-\Delta_{\alpha\beta,k}/\omega_c}\,  {\rm Ei}(\Delta_{\alpha\beta,k}/\omega_c)
\label{eq:Iminus}
\end{equation}
with ${\rm Ei}(z)=-\int_{-z}^\infty dy\, {\rm e}^{-y}/y$, where the integral is understood in the principal value sense. The terms linear in $\omega_c$ in \eqref{eq:Iminus} describing the static effect of the bath do not contribute to the Lamb shift \cite{santana:2001} as they cancel each other. Finally, the Lamb shift (\ref{eq:lscompact}) reduces for a TLS in the zero temperature limit to
\begin{eqnarray}\label{lamb1}
\delta\omega_{\alpha\beta}^{T=0}=\frac{\eta}{\pi}\sum_{\mu, k}\bigl[\Delta_{\beta\mu,k} f(\Delta_{\beta\mu,k}/\omega_c)|X_{\beta\mu,k}^{(r)}|^2
- \Delta_{\alpha\mu,-k} f(\Delta_{\alpha\mu,-k}/\omega_c)|X_{\mu\alpha,k}^{(r)}|^2\bigr]\,,
\end{eqnarray}
where we have introduced the function $f(x)={\rm Ei(x)} {\rm exp}(-x)$ with the property $f(x) \to C_{\gamma} + \ln(x)$ for $ x \ll 1$ with the Euler constant $C_{\gamma}$. For the Rabi model discussed in the main text, this yields the expressions in (5) together with $\Lambda^{(r)}=\sum_k g(\Delta_{21,k}/\omega_c) |X_{21,k}^{(r)}|^2$, where $g(x)=x\,[f(x)+ f(-x)]$.

Thermal corrections to these zero temperature results tend to play a role for dressed transition frequencies with $\hbar\beta\Delta_{12, k} \lesssim O(1)$, while they are suppressed exponentially for $\hbar\beta\Delta_{12, k}\gg 1$.
We discuss details for the Rabi model below. In principle, there is also a constraint on temperature corresponding to the Markov-approximation associated with (\ref{eq:standardnotation}), namely, that bath correlation functions decay sufficiently fast compared to a typical system relaxation time scale $1/\Gamma$. However, in a steady state situation, the time-independence of the density guarantees that non-Markovian effects are of no relevance. They may only play a role when one is interested in time-dependent correlation functions.

\section{Semiclassical Rabi model}

We consider here details of the semiclassical Rabi model which, despite its simplicity, describes recent experimental realizations, see below and, e.g., \cite{Oliver1:2005, Wilson1:2007, Wilson1:2010}. The Hamiltonian is given by
\begin{equation}
H(t)=\twobytwo{-\frac{E}{2}}{A^*\,{\rm e}^{i\Omega t}}{A\,{\rm e}^{-i\Omega t}}{\frac{E}{2}}=-\frac{E}{2}\sigma_z + A\,\{\cos(\Omega t)\sigma_x - \sin(\Omega t) \sigma_y\}\,,
\label{eq:Rabihamiltonian1}
\end{equation}
where for simplicity the driving amplitude $A$ is taken as real-valued and $E$ is the bare energy level spacing.
To find the solution of the Schr\"odinger equation, we follow a standard procedure: First, moving to a rotating frame reveals a time-independent Hamiltonian which we diagonalize. In a second step, we revert to the laboratory frame and cast the solutions finally in Floquet form. The so-found quasienergies are $\epsilon_{1,2}=(\Delta\pm\hbar\omega_R)/2$ with detuning $\Delta=E-\hbar\Omega$ and $\omega_R=\frac{1}{\hbar}\sqrt{\Delta^2 + 4|A|^2}$ being the Rabi frequency.\\
The corresponding Floquet modes are:
\begin{equation}
\ket{\phi_1(t)}= \begin{pmatrix} \cos \theta \\ -\,{\rm e}^{-i\Omega t}\,{\rm e}^{i\phi}\sin\theta \end{pmatrix}
\label{eq:Floqmode1}
\end{equation}
\begin{equation}
\ket{\phi_2(t)}= \begin{pmatrix} {\rm e}^{-i\phi}\sin\theta \\ \,{\rm e}^{-i\Omega t} \cos\theta \end{pmatrix}
\label{eq:Floqmode2}
\end{equation}
where $\phi=-\arg A$ and $\tan(2\theta)=\frac{2|A|}\Delta$.

As we have chosen $A$ to be real, one has $\phi=0$.
The coupling matrix elements $X_{\alpha \beta, k}^{(r)}$ are obtained in the following way:
 For $\sigma_z$-noise only terms with $k=0$ survive with
\begin{align}\label{mat1}
X_{11,0}^{(0)} &= \cos^2 \theta - \sin^2\theta = \cos 2 \theta =-X_{22,0}^{(0)}\nonumber\\
X_{21,0}^{(0)} &= 2\sin\theta \cos \theta = \sin 2\theta = X_{12,0}^{\ast \;(0)}\,.
\end{align}
Together with $\Lambda^{(r)}$ specified in the preceding section, this then leads to $\Lambda^{(0)}$.
For $\sigma_x$-noise, the non-zero contributions $k=\pm1$ provide the coefficients
\begin{align}\label{mat2}
X_{11,-1}^{(\pi/2)} &= -\frac{1}{2} \sin 2\theta = X_{11,1}^{(\pi/2)}\ ,
&X_{22,-1}^{(\pi/2)} &= X_{22,1}^{(\pi/2)} = -X_{11,1}^{(\pi/2)} \nonumber\\
X_{12,-1}^{(\pi/2)} &= \cos^2\theta = X_{21,1}^{\ast\;(\pi/2)}\ ,
&X_{21,-1}^{(\pi/2)} &= -\sin^2\theta = X_{12,1}^{\ast\;(\pi/2)}\ ,
\end{align}
 which yields $\Lambda^{(\pi/2)}$ [Eq. (5) in the main text].

 Thermal fluctuations in the master equation (\ref{eq:standardnotation}) and the Lamb shift (\ref{lamb1}) are important if
 $(-\omega_R\pm \Omega)\hbar\beta \lesssim 1$ ($\sigma_x$-coupling) or $\omega_R\hbar\beta\lesssim 1$ ($\sigma_z$-coupling).
 In actual experiments \cite{Oliver1:2005, Wilson1:2007, Wilson1:2010}, close to resonance one typically has  $\Omega\gg \eta |A|$ so that thermal fluctuations can be sufficiently suppressed for $\sigma_x$-noise, where this coincides with the optimal regime to detect a Lamb shift  (see main text). For $\sigma_z$-coupling the constraint is in general harder to fulfill. However, since a corresponding Lamb shift close to resonance is particularly pronounced in the range of stronger driving ($|A|/E\gtrsim 0.3$), for realistic values $E\beta\approx 10$ the condition $\omega_R\hbar\beta\approx (2|A|/E) (E\beta)\approx 6$ is sufficiently obeyed as well.

 \section{Single mode case}
 While it makes in general no sense to derive a master equation such as in (\ref{eq:standardnotation}) for a single mode reservoir, the result derived in Eq.~(4) of the main text for the Lamb shift  can also be applied to analyze this limiting case. It is only based on zero temperature second order perturbation theory. To illustrate the enhanced Lamb shift for dressed states we thus consider here a TLS interacting with such a single mode reservoir with frequency $\omega_0$. With the distribution $J_1(\omega)=\eta\hbar\omega_0^2\,  \delta(\omega-\omega_0)$ the integrals (\ref{eq:PVintegral}) are easily evaluated and based on the coupling matrix elements (\ref{mat1}), (\ref{mat2}), one obtains for  $\sigma_z$-noise
\begin{equation}
\delta\omega_{12}^{\rm (single, z)}= \frac{\eta}{\pi} \,  2\omega_R \sin(2\theta)\, \frac{ \omega_0^2}{\omega_0^2-\omega_R^2}\,.
\end{equation}
For $\sigma_x$-noise, we have
\begin{equation}
\delta\omega_{12}^{\rm (single, x)}=\frac{\eta\omega_0^2}{\pi}\left[\frac{(\omega_R+\Omega)\cos^4(\theta)}{\omega_0^2-(\omega_R+\Omega)^2}
+\frac{\Omega\sin^2(\theta)/4}{\omega_0^2-\Omega^2}\right]\, .
\end{equation}
In the static situation (no driving) the so-called dispersive limit ($E$ and $\omega_0$ far detuned) has been considered in \cite{Fragner1:2008}. It yields
\begin{equation}
\delta\omega_{12}^{(\rm static)}\approx \frac{\eta}{\pi} \frac{\omega_0^2}{(E/\hbar)-\omega_0}\, .
\end{equation}
While this static Lamb shift leads to shifts in bare transition frequencies of a few \% only, apparently, by properly tuning $A$ and $\Omega$ the dressed Lamb shift can be significantly enhanced.

\section{Mapping to the Rabi model}

As discussed in the main text, recent experiments with driven CPBs are described with
\begin{equation}\label{hamcpb}
H_{S, \rm CPB}(t)=-\frac{1}{2} E_{C}\tau_z -\frac{1}{2} E_J \tau_x - \lambda \cos(\omega t) \tau_z
\end{equation}
with tunable charging energy $E_C$ and Josephson energy $E_J$ \cite{Oliver1:2005, Wilson1:2007, Wilson1:2010}. The coupling to the reservoir is dominated by charge noise and thus proportional to $\tau_z$. Here, we show  how in limiting cases this setup reduces to the Rabi model (\ref{eq:Rabihamiltonian1}) with either transversal or longitudinal coupling to the bath.

Transversal coupling: Close to charge degeneracy $E_C=0$, in the eigenstate representation of the CPB, i.e.,\ $\tau_x\to \sigma_z$ and $\tau_z\to -\sigma_x$, one has $E=E_J$ and the coupling to the bath is transversal with $S(r=\pi/2)=\sigma_x$. Close to resonance $\hbar\omega\approx E_J$ a RWA in $H_{\rm CPB}$  can be applied which is consistent with the PSA used to derive the Floquet master equation (\ref{eq:standardnotation}) if $\hbar\omega/|\lambda|\gg 1\gg \eta$. The drive parameters then read $A=\lambda$ and $\Omega=-\omega$.

Longitudinal coupling: For $E_C\neq 0$  and for strong driving $\lambda, \hbar\omega\gg E_J$, a dressed tunneling picture applies. Accordingly, a unitary transformation $U(t)=\exp[-i\phi(t)\tau_z/(2\hbar)]$ with $\dot{\phi}(t)=2\lambda \cos(\omega t)$ leads to
\begin{eqnarray}
\tilde{H}_{S,\rm CPB}(t) &=&U^\dagger H_{S, \rm CPB}(t) U(t)+i\hbar U^\dagger(t)\dot{U}(t)\nonumber\\
&=&-\frac{1}{2} E_{C}\tau_z -\frac{1}{2} E_{J}\left[ {\rm e}^{i\phi(t)/\hbar}\tau_+ + {\rm e}^{-i\phi(t)/\hbar}\tau_-\right]
\end{eqnarray}
with $\tau_\pm=(\tau_x\pm i\tau_y)/2$. By decomposing the time-dependent phase factors in terms of Bessel functions using the Jacobi-Anger expansion $\exp(i z \sin \varphi)=\sum_{-\infty}^{\infty} J_n(z) \exp(in\varphi)$,  near the $n$-photon resonance $E_C=n\hbar\omega$ one may put ${\rm e}^{i\phi(t)/\hbar}\approx J_n(2\lambda/\hbar \omega)\, {\rm e}^{in\omega t}$.
This RWA is consistent with the PSA provided $\hbar\omega/E_J\gg 1$ as assumed.

Accordingly, one obtains the Rabi model with $E=E_C$ and drive parameters $A=-(E_J/2)\,  J_n(2\lambda/\hbar\omega), \Omega=n\omega$. As the transformation $U(t)$ commutes with the CPB-bath coupling operator $\tau_z$, one arrives at pure dephasing (longitudinal coupling).

\bibliographystyle{apsrev4-1}

\end{widetext}

\end{document}